\documentstyle[sprocl]{article}

\def\Journal#1#2#3#4{{#1} {\bf #2}, #3 (#4)}

\def\NCA{\em Nuovo Cimento}
\def\NIM{\em Nucl. Instrum. Methods}
\def\NIMA{{\em Nucl. Instrum. Methods} A}
\def\NPB{{\em Nucl. Phys.} B}
\def\PLB{{\em Phys. Lett.}  B}
\def\PRL{\em Phys. Rev. Lett.}
\def\PRB{{\em Phys. Rev.} B}
\def\SSC{\em Solid State Commun.}
\def\ZPC{{\em Z. Phys.} C}
\def\st{\scriptstyle}
\def\sst{\scriptscriptstyle}
\def\mco{\multicolumn}
\def\epp{\epsilon^{\prime}}
\def\vep{\varepsilon}
\def\ra{\rightarrow}
\def\ppg{\pi^+\pi^-\gamma}
\def\vp{{\bf p}}
\def\ko{K^0}
\def\kb{\bar{K^0}}
\def\al{\alpha}
\def\ab{\bar{\alpha}}
\def\be{\begin{equation}}
\def\ee{\end{equation}}
\def\bea{\begin{eqnarray}}
\def\eea{\end{eqnarray}}
\def\CPbar{\hbox{{\rm CP}\hskip-1.80em{/}}}

\begin{document}

\title{COLLECTIVE ROTATIONS IN OXIDES}

\author{ EMILIO ARTACHO }

\address{Departamento de F\'{\i}sica de la Materia Condensada, C-III \\
Universidad Aut\'onoma de Madrid, 28049 Madrid, Spain}

\author{ KANWAL G. SINGH }

\address{Department of Physics, Wellesley College, Wellesley, MA 02181, USA}

\author{ GUILLERMO GOMEZ-SANTOS }

\address{Departamento de F\'{\i}sica de la Materia Condensada, C-III \\
Universidad Aut\'onoma de Madrid, 28049 Madrid, Spain}

\maketitle\abstracts{
Based on bond arguments, a Hamiltonian is introduced to describe the 
fundamental physics of the collective rotations of oxygen atoms in oxides.
Values for the relevant material parameters are estimated for silica 
and Cu oxides.
Zero-temperature phase diagrams are presented; the different phases and
associated transitions are characterized. Phases with non-phononic 
excitations are found.}

\section{Introduction}

\vspace{-3pt}

The tendency of oxygen atoms to form just two bonds with a bond angle 
smaller than 180$^{\rm o}$ allows the appearance of low energy atomic motions
related to internal rotational degrees of freedom which, in a first
approximation, involve neither bending nor stretching of bonds 
(``rigid unit modes" in silicates, ``floppy modes" in glasses).\cite{beta}
A huge variety of compounds exhibit the oxygen-bridge $X$-O-$X$ basic unit,
$X$ being a linking atom. Among the many silicates, $\beta$-cristobalite
(a silica polymorph) shows these internal rotations most clearly.\cite{beta}
The perovskites are also interesting candidate materials, especially the
CuO$_2$ planes of the high $T_c$ superconducting materials. 

\begin{figure}
\begin{center}
\mbox{ \hspace {-1.cm} 
\setlength{\unitlength}{0.240900pt}
\ifx\plotpoint\undefined\newsavebox{\plotpoint}\fi
\sbox{\plotpoint}{\rule[-0.175pt]{0.350pt}{0.350pt}}%
\begin{picture}(779,600)(0,0)
\tenrm
\sbox{\plotpoint}{\rule[-0.175pt]{0.350pt}{0.350pt}}%
\put(264,255){\rule[-0.175pt]{4.818pt}{0.350pt}}
\put(242,255){\makebox(0,0)[r]{\small $5$}}
\put(695,255){\rule[-0.175pt]{4.818pt}{0.350pt}}
\put(264,352){\rule[-0.175pt]{4.818pt}{0.350pt}}
\put(242,352){\makebox(0,0)[r]{\small $10$}}
\put(695,352){\rule[-0.175pt]{4.818pt}{0.350pt}}
\put(264,448){\rule[-0.175pt]{4.818pt}{0.350pt}}
\put(242,448){\makebox(0,0)[r]{\small $15$}}
\put(695,448){\rule[-0.175pt]{4.818pt}{0.350pt}}
\put(397,158){\rule[-0.175pt]{0.350pt}{4.818pt}}
\put(397,113){\makebox(0,0){\small $5$}}
\put(397,467){\rule[-0.175pt]{0.350pt}{4.818pt}}
\put(529,158){\rule[-0.175pt]{0.350pt}{4.818pt}}
\put(529,113){\makebox(0,0){\small $10$}}
\put(529,467){\rule[-0.175pt]{0.350pt}{4.818pt}}
\put(662,158){\rule[-0.175pt]{0.350pt}{4.818pt}}
\put(662,113){\makebox(0,0){\small $15$}}
\put(662,467){\rule[-0.175pt]{0.350pt}{4.818pt}}
\put(264,158){\rule[-0.175pt]{108.646pt}{0.350pt}}
\put(715,158){\rule[-0.175pt]{0.350pt}{79.256pt}}
\put(264,487){\rule[-0.175pt]{108.646pt}{0.350pt}}
\put(89,322){\makebox(0,0)[l]{\shortstack{\large $B$,$t$ \\ (meV)}}}
\put(533,68){\makebox(0,0){\Large $\alpha$ ($^{\rm o}$)}}
\put(397,429){\makebox(0,0)[l]{\large $B$}}
\put(609,429){\makebox(0,0)[l]{\Large $t$}}
\put(264,158){\rule[-0.175pt]{0.350pt}{79.256pt}}
\sbox{\plotpoint}{\rule[-0.500pt]{1.000pt}{1.000pt}}%
\put(264,158){\usebox{\plotpoint}}
\put(264,158){\rule[-0.500pt]{3.373pt}{1.000pt}}
\put(278,159){\rule[-0.500pt]{3.132pt}{1.000pt}}
\put(291,160){\rule[-0.500pt]{1.204pt}{1.000pt}}
\put(296,161){\usebox{\plotpoint}}
\put(300,162){\rule[-0.500pt]{1.204pt}{1.000pt}}
\put(305,163){\rule[-0.500pt]{1.204pt}{1.000pt}}
\put(310,164){\usebox{\plotpoint}}
\put(314,165){\usebox{\plotpoint}}
\put(316,166){\usebox{\plotpoint}}
\put(319,167){\usebox{\plotpoint}}
\put(323,168){\usebox{\plotpoint}}
\put(325,169){\usebox{\plotpoint}}
\put(328,170){\usebox{\plotpoint}}
\put(332,171){\usebox{\plotpoint}}
\put(334,172){\usebox{\plotpoint}}
\put(337,173){\usebox{\plotpoint}}
\put(339,174){\usebox{\plotpoint}}
\put(341,175){\usebox{\plotpoint}}
\put(343,176){\usebox{\plotpoint}}
\put(346,177){\usebox{\plotpoint}}
\put(348,178){\usebox{\plotpoint}}
\put(351,179){\usebox{\plotpoint}}
\put(352,180){\usebox{\plotpoint}}
\put(353,181){\usebox{\plotpoint}}
\put(355,182){\usebox{\plotpoint}}
\put(357,183){\usebox{\plotpoint}}
\put(360,184){\usebox{\plotpoint}}
\put(361,185){\usebox{\plotpoint}}
\put(362,186){\usebox{\plotpoint}}
\put(364,187){\usebox{\plotpoint}}
\put(366,188){\usebox{\plotpoint}}
\put(369,189){\usebox{\plotpoint}}
\put(370,190){\usebox{\plotpoint}}
\put(371,191){\usebox{\plotpoint}}
\put(373,192){\usebox{\plotpoint}}
\put(374,193){\usebox{\plotpoint}}
\put(376,194){\usebox{\plotpoint}}
\put(377,195){\usebox{\plotpoint}}
\put(379,196){\usebox{\plotpoint}}
\put(380,197){\usebox{\plotpoint}}
\put(382,198){\usebox{\plotpoint}}
\put(383,199){\usebox{\plotpoint}}
\put(385,200){\usebox{\plotpoint}}
\put(386,201){\usebox{\plotpoint}}
\put(388,202){\usebox{\plotpoint}}
\put(390,203){\usebox{\plotpoint}}
\put(391,204){\usebox{\plotpoint}}
\put(393,205){\usebox{\plotpoint}}
\put(394,206){\usebox{\plotpoint}}
\put(395,207){\usebox{\plotpoint}}
\put(396,208){\usebox{\plotpoint}}
\put(397,209){\usebox{\plotpoint}}
\put(399,210){\usebox{\plotpoint}}
\put(400,211){\usebox{\plotpoint}}
\put(402,212){\usebox{\plotpoint}}
\put(403,213){\usebox{\plotpoint}}
\put(404,214){\usebox{\plotpoint}}
\put(405,215){\usebox{\plotpoint}}
\put(406,216){\usebox{\plotpoint}}
\put(408,217){\usebox{\plotpoint}}
\put(409,218){\usebox{\plotpoint}}
\put(411,219){\usebox{\plotpoint}}
\put(412,220){\usebox{\plotpoint}}
\put(413,221){\usebox{\plotpoint}}
\put(414,222){\usebox{\plotpoint}}
\put(415,223){\usebox{\plotpoint}}
\put(416,224){\usebox{\plotpoint}}
\put(417,225){\usebox{\plotpoint}}
\put(419,226){\usebox{\plotpoint}}
\put(420,227){\usebox{\plotpoint}}
\put(421,228){\usebox{\plotpoint}}
\put(422,229){\usebox{\plotpoint}}
\put(423,230){\usebox{\plotpoint}}
\put(424,231){\usebox{\plotpoint}}
\put(425,232){\usebox{\plotpoint}}
\put(426,233){\usebox{\plotpoint}}
\put(428,234){\usebox{\plotpoint}}
\put(429,235){\usebox{\plotpoint}}
\put(430,236){\usebox{\plotpoint}}
\put(431,237){\usebox{\plotpoint}}
\put(432,238){\usebox{\plotpoint}}
\put(433,239){\usebox{\plotpoint}}
\put(434,240){\usebox{\plotpoint}}
\put(435,241){\usebox{\plotpoint}}
\put(436,242){\usebox{\plotpoint}}
\put(437,243){\usebox{\plotpoint}}
\put(438,244){\usebox{\plotpoint}}
\put(439,245){\usebox{\plotpoint}}
\put(440,246){\usebox{\plotpoint}}
\put(441,247){\usebox{\plotpoint}}
\put(442,248){\usebox{\plotpoint}}
\put(443,249){\usebox{\plotpoint}}
\put(444,250){\usebox{\plotpoint}}
\put(445,251){\usebox{\plotpoint}}
\put(446,252){\usebox{\plotpoint}}
\put(447,253){\usebox{\plotpoint}}
\put(448,254){\usebox{\plotpoint}}
\put(449,255){\usebox{\plotpoint}}
\put(450,256){\usebox{\plotpoint}}
\put(451,257){\usebox{\plotpoint}}
\put(452,258){\usebox{\plotpoint}}
\put(453,259){\usebox{\plotpoint}}
\put(454,260){\usebox{\plotpoint}}
\put(455,262){\usebox{\plotpoint}}
\put(456,263){\usebox{\plotpoint}}
\put(457,264){\usebox{\plotpoint}}
\put(458,265){\usebox{\plotpoint}}
\put(459,266){\usebox{\plotpoint}}
\put(460,267){\usebox{\plotpoint}}
\put(461,268){\usebox{\plotpoint}}
\put(462,269){\usebox{\plotpoint}}
\put(463,270){\usebox{\plotpoint}}
\put(464,272){\usebox{\plotpoint}}
\put(465,273){\usebox{\plotpoint}}
\put(466,274){\usebox{\plotpoint}}
\put(467,275){\usebox{\plotpoint}}
\put(468,276){\usebox{\plotpoint}}
\put(469,277){\usebox{\plotpoint}}
\put(470,278){\usebox{\plotpoint}}
\put(471,279){\usebox{\plotpoint}}
\put(472,280){\usebox{\plotpoint}}
\put(473,281){\usebox{\plotpoint}}
\put(474,282){\usebox{\plotpoint}}
\put(475,283){\usebox{\plotpoint}}
\put(476,285){\usebox{\plotpoint}}
\put(477,286){\usebox{\plotpoint}}
\put(478,288){\usebox{\plotpoint}}
\put(479,289){\usebox{\plotpoint}}
\put(480,290){\usebox{\plotpoint}}
\put(481,291){\usebox{\plotpoint}}
\put(482,292){\usebox{\plotpoint}}
\put(483,293){\usebox{\plotpoint}}
\put(484,294){\usebox{\plotpoint}}
\put(485,296){\usebox{\plotpoint}}
\put(486,297){\usebox{\plotpoint}}
\put(487,299){\usebox{\plotpoint}}
\put(488,300){\usebox{\plotpoint}}
\put(489,301){\usebox{\plotpoint}}
\put(490,302){\usebox{\plotpoint}}
\put(491,303){\usebox{\plotpoint}}
\put(492,305){\usebox{\plotpoint}}
\put(493,306){\usebox{\plotpoint}}
\put(494,308){\usebox{\plotpoint}}
\put(495,309){\usebox{\plotpoint}}
\put(496,311){\usebox{\plotpoint}}
\put(497,312){\usebox{\plotpoint}}
\put(498,313){\usebox{\plotpoint}}
\put(499,314){\usebox{\plotpoint}}
\put(500,315){\usebox{\plotpoint}}
\put(501,317){\usebox{\plotpoint}}
\put(502,318){\usebox{\plotpoint}}
\put(503,320){\usebox{\plotpoint}}
\put(504,321){\usebox{\plotpoint}}
\put(505,323){\usebox{\plotpoint}}
\put(506,324){\usebox{\plotpoint}}
\put(507,325){\usebox{\plotpoint}}
\put(508,326){\usebox{\plotpoint}}
\put(509,327){\usebox{\plotpoint}}
\put(510,329){\usebox{\plotpoint}}
\put(511,330){\usebox{\plotpoint}}
\put(512,331){\usebox{\plotpoint}}
\put(513,332){\usebox{\plotpoint}}
\put(514,333){\usebox{\plotpoint}}
\put(515,335){\usebox{\plotpoint}}
\put(516,336){\usebox{\plotpoint}}
\put(517,338){\usebox{\plotpoint}}
\put(518,340){\usebox{\plotpoint}}
\put(519,342){\usebox{\plotpoint}}
\put(520,343){\usebox{\plotpoint}}
\put(521,344){\usebox{\plotpoint}}
\put(522,345){\usebox{\plotpoint}}
\put(523,346){\usebox{\plotpoint}}
\put(524,348){\usebox{\plotpoint}}
\put(525,349){\usebox{\plotpoint}}
\put(526,351){\usebox{\plotpoint}}
\put(527,353){\usebox{\plotpoint}}
\put(528,355){\usebox{\plotpoint}}
\put(529,356){\usebox{\plotpoint}}
\put(530,357){\usebox{\plotpoint}}
\put(531,359){\usebox{\plotpoint}}
\put(532,360){\usebox{\plotpoint}}
\put(533,361){\usebox{\plotpoint}}
\put(534,363){\usebox{\plotpoint}}
\put(535,365){\usebox{\plotpoint}}
\put(536,367){\usebox{\plotpoint}}
\put(537,369){\usebox{\plotpoint}}
\put(538,370){\usebox{\plotpoint}}
\put(539,371){\usebox{\plotpoint}}
\put(540,373){\usebox{\plotpoint}}
\put(541,374){\usebox{\plotpoint}}
\put(542,375){\usebox{\plotpoint}}
\put(543,377){\usebox{\plotpoint}}
\put(544,379){\usebox{\plotpoint}}
\put(545,381){\usebox{\plotpoint}}
\put(546,383){\usebox{\plotpoint}}
\put(547,384){\usebox{\plotpoint}}
\put(548,385){\usebox{\plotpoint}}
\put(549,387){\usebox{\plotpoint}}
\put(550,388){\usebox{\plotpoint}}
\put(551,389){\usebox{\plotpoint}}
\put(552,391){\usebox{\plotpoint}}
\put(553,392){\usebox{\plotpoint}}
\put(554,394){\usebox{\plotpoint}}
\put(555,395){\usebox{\plotpoint}}
\put(556,396){\usebox{\plotpoint}}
\put(557,399){\usebox{\plotpoint}}
\put(558,401){\usebox{\plotpoint}}
\put(559,403){\usebox{\plotpoint}}
\put(560,405){\usebox{\plotpoint}}
\put(561,406){\usebox{\plotpoint}}
\put(562,407){\usebox{\plotpoint}}
\put(563,409){\usebox{\plotpoint}}
\put(564,410){\usebox{\plotpoint}}
\put(565,411){\usebox{\plotpoint}}
\put(566,414){\usebox{\plotpoint}}
\put(567,416){\usebox{\plotpoint}}
\put(568,418){\usebox{\plotpoint}}
\put(569,420){\usebox{\plotpoint}}
\put(570,421){\usebox{\plotpoint}}
\put(571,423){\usebox{\plotpoint}}
\put(572,424){\usebox{\plotpoint}}
\put(573,426){\usebox{\plotpoint}}
\put(574,428){\usebox{\plotpoint}}
\put(575,429){\usebox{\plotpoint}}
\put(576,431){\usebox{\plotpoint}}
\put(577,433){\usebox{\plotpoint}}
\put(578,435){\usebox{\plotpoint}}
\put(579,436){\usebox{\plotpoint}}
\put(580,438){\usebox{\plotpoint}}
\put(581,439){\usebox{\plotpoint}}
\put(582,441){\usebox{\plotpoint}}
\put(583,443){\usebox{\plotpoint}}
\put(584,445){\usebox{\plotpoint}}
\put(585,447){\usebox{\plotpoint}}
\put(586,449){\usebox{\plotpoint}}
\put(587,452){\usebox{\plotpoint}}
\put(588,453){\usebox{\plotpoint}}
\put(589,455){\usebox{\plotpoint}}
\put(590,456){\usebox{\plotpoint}}
\put(591,458){\usebox{\plotpoint}}
\put(592,460){\usebox{\plotpoint}}
\put(593,461){\usebox{\plotpoint}}
\put(594,463){\usebox{\plotpoint}}
\put(595,464){\usebox{\plotpoint}}
\put(596,466){\usebox{\plotpoint}}
\put(597,468){\usebox{\plotpoint}}
\put(598,470){\usebox{\plotpoint}}
\put(599,472){\usebox{\plotpoint}}
\put(600,474){\usebox{\plotpoint}}
\put(601,476){\usebox{\plotpoint}}
\put(602,477){\usebox{\plotpoint}}
\put(603,479){\usebox{\plotpoint}}
\put(604,481){\usebox{\plotpoint}}
\put(605,483){\usebox{\plotpoint}}
\put(606,484){\usebox{\plotpoint}}
\put(354,481){\rule[-0.500pt]{1.000pt}{1.445pt}}
\put(355,475){\rule[-0.500pt]{1.000pt}{1.445pt}}
\put(356,469){\rule[-0.500pt]{1.000pt}{1.445pt}}
\put(357,463){\rule[-0.500pt]{1.000pt}{1.445pt}}
\put(358,457){\rule[-0.500pt]{1.000pt}{1.445pt}}
\put(359,451){\rule[-0.500pt]{1.000pt}{1.445pt}}
\put(360,444){\rule[-0.500pt]{1.000pt}{1.566pt}}
\put(361,438){\rule[-0.500pt]{1.000pt}{1.566pt}}
\put(362,431){\rule[-0.500pt]{1.000pt}{1.566pt}}
\put(363,425){\rule[-0.500pt]{1.000pt}{1.566pt}}
\put(364,420){\rule[-0.500pt]{1.000pt}{1.060pt}}
\put(365,416){\rule[-0.500pt]{1.000pt}{1.060pt}}
\put(366,411){\rule[-0.500pt]{1.000pt}{1.060pt}}
\put(367,407){\rule[-0.500pt]{1.000pt}{1.060pt}}
\put(368,403){\rule[-0.500pt]{1.000pt}{1.060pt}}
\put(369,398){\rule[-0.500pt]{1.000pt}{1.205pt}}
\put(370,393){\rule[-0.500pt]{1.000pt}{1.204pt}}
\put(371,388){\rule[-0.500pt]{1.000pt}{1.204pt}}
\put(372,383){\rule[-0.500pt]{1.000pt}{1.204pt}}
\put(373,379){\usebox{\plotpoint}}
\put(374,375){\usebox{\plotpoint}}
\put(375,372){\usebox{\plotpoint}}
\put(376,368){\usebox{\plotpoint}}
\put(377,365){\usebox{\plotpoint}}
\put(378,361){\usebox{\plotpoint}}
\put(379,357){\usebox{\plotpoint}}
\put(380,353){\usebox{\plotpoint}}
\put(381,350){\usebox{\plotpoint}}
\put(382,347){\usebox{\plotpoint}}
\put(383,344){\usebox{\plotpoint}}
\put(384,341){\usebox{\plotpoint}}
\put(385,338){\usebox{\plotpoint}}
\put(386,336){\usebox{\plotpoint}}
\put(387,333){\usebox{\plotpoint}}
\put(388,330){\usebox{\plotpoint}}
\put(389,328){\usebox{\plotpoint}}
\put(390,325){\usebox{\plotpoint}}
\put(391,323){\usebox{\plotpoint}}
\put(392,320){\usebox{\plotpoint}}
\put(393,317){\usebox{\plotpoint}}
\put(394,314){\usebox{\plotpoint}}
\put(395,312){\usebox{\plotpoint}}
\put(396,310){\usebox{\plotpoint}}
\put(397,308){\usebox{\plotpoint}}
\put(398,306){\usebox{\plotpoint}}
\put(399,304){\usebox{\plotpoint}}
\put(400,302){\usebox{\plotpoint}}
\put(401,299){\usebox{\plotpoint}}
\put(402,297){\usebox{\plotpoint}}
\put(403,295){\usebox{\plotpoint}}
\put(404,293){\usebox{\plotpoint}}
\put(405,291){\usebox{\plotpoint}}
\put(406,289){\usebox{\plotpoint}}
\put(407,288){\usebox{\plotpoint}}
\put(408,286){\usebox{\plotpoint}}
\put(409,285){\usebox{\plotpoint}}
\put(410,283){\usebox{\plotpoint}}
\put(411,281){\usebox{\plotpoint}}
\put(412,279){\usebox{\plotpoint}}
\put(413,277){\usebox{\plotpoint}}
\put(414,275){\usebox{\plotpoint}}
\put(415,274){\usebox{\plotpoint}}
\put(416,272){\usebox{\plotpoint}}
\put(417,271){\usebox{\plotpoint}}
\put(418,270){\usebox{\plotpoint}}
\put(419,268){\usebox{\plotpoint}}
\put(420,267){\usebox{\plotpoint}}
\put(421,265){\usebox{\plotpoint}}
\put(422,264){\usebox{\plotpoint}}
\put(423,262){\usebox{\plotpoint}}
\put(424,261){\usebox{\plotpoint}}
\put(425,260){\usebox{\plotpoint}}
\put(426,259){\usebox{\plotpoint}}
\put(427,258){\usebox{\plotpoint}}
\put(428,258){\usebox{\plotpoint}}
\put(429,257){\usebox{\plotpoint}}
\put(430,256){\usebox{\plotpoint}}
\put(431,255){\usebox{\plotpoint}}
\put(432,254){\usebox{\plotpoint}}
\put(433,251){\usebox{\plotpoint}}
\put(434,250){\usebox{\plotpoint}}
\put(435,249){\usebox{\plotpoint}}
\put(436,248){\usebox{\plotpoint}}
\put(437,248){\usebox{\plotpoint}}
\put(438,247){\usebox{\plotpoint}}
\put(439,246){\usebox{\plotpoint}}
\put(440,245){\usebox{\plotpoint}}
\put(441,244){\usebox{\plotpoint}}
\put(442,243){\usebox{\plotpoint}}
\put(443,242){\usebox{\plotpoint}}
\put(444,241){\usebox{\plotpoint}}
\put(445,240){\usebox{\plotpoint}}
\put(446,239){\usebox{\plotpoint}}
\put(447,238){\usebox{\plotpoint}}
\put(448,237){\usebox{\plotpoint}}
\put(449,236){\usebox{\plotpoint}}
\put(451,235){\usebox{\plotpoint}}
\put(452,234){\usebox{\plotpoint}}
\put(453,233){\usebox{\plotpoint}}
\put(455,232){\usebox{\plotpoint}}
\put(456,231){\usebox{\plotpoint}}
\put(457,230){\usebox{\plotpoint}}
\put(458,229){\usebox{\plotpoint}}
\put(460,228){\usebox{\plotpoint}}
\put(461,227){\usebox{\plotpoint}}
\put(462,226){\usebox{\plotpoint}}
\put(464,225){\usebox{\plotpoint}}
\put(465,224){\usebox{\plotpoint}}
\put(467,223){\usebox{\plotpoint}}
\put(468,222){\usebox{\plotpoint}}
\put(470,221){\usebox{\plotpoint}}
\put(472,220){\usebox{\plotpoint}}
\put(473,219){\usebox{\plotpoint}}
\put(476,218){\usebox{\plotpoint}}
\put(478,217){\usebox{\plotpoint}}
\put(479,216){\usebox{\plotpoint}}
\put(481,215){\usebox{\plotpoint}}
\put(482,214){\usebox{\plotpoint}}
\put(485,213){\usebox{\plotpoint}}
\put(487,212){\usebox{\plotpoint}}
\put(489,211){\usebox{\plotpoint}}
\put(492,210){\usebox{\plotpoint}}
\put(494,209){\usebox{\plotpoint}}
\put(496,208){\usebox{\plotpoint}}
\put(498,207){\usebox{\plotpoint}}
\put(501,206){\usebox{\plotpoint}}
\put(503,205){\usebox{\plotpoint}}
\put(505,204){\rule[-0.500pt]{1.204pt}{1.000pt}}
\put(510,203){\usebox{\plotpoint}}
\put(512,202){\usebox{\plotpoint}}
\put(515,201){\usebox{\plotpoint}}
\put(519,200){\usebox{\plotpoint}}
\put(521,199){\usebox{\plotpoint}}
\put(524,198){\usebox{\plotpoint}}
\put(528,197){\rule[-0.500pt]{1.204pt}{1.000pt}}
\put(533,196){\usebox{\plotpoint}}
\put(535,195){\usebox{\plotpoint}}
\put(537,194){\rule[-0.500pt]{1.204pt}{1.000pt}}
\put(542,193){\usebox{\plotpoint}}
\put(546,192){\rule[-0.500pt]{1.204pt}{1.000pt}}
\put(551,191){\rule[-0.500pt]{1.204pt}{1.000pt}}
\put(556,190){\usebox{\plotpoint}}
\put(560,189){\rule[-0.500pt]{1.204pt}{1.000pt}}
\put(565,188){\usebox{\plotpoint}}
\put(569,187){\rule[-0.500pt]{1.204pt}{1.000pt}}
\put(574,186){\rule[-0.500pt]{2.168pt}{1.000pt}}
\put(583,185){\usebox{\plotpoint}}
\put(587,184){\rule[-0.500pt]{1.204pt}{1.000pt}}
\put(592,183){\rule[-0.500pt]{2.168pt}{1.000pt}}
\put(601,182){\rule[-0.500pt]{1.204pt}{1.000pt}}
\put(606,181){\rule[-0.500pt]{2.168pt}{1.000pt}}
\put(615,180){\rule[-0.500pt]{2.168pt}{1.000pt}}
\put(624,179){\rule[-0.500pt]{2.168pt}{1.000pt}}
\put(633,178){\rule[-0.500pt]{2.168pt}{1.000pt}}
\put(642,177){\rule[-0.500pt]{2.168pt}{1.000pt}}
\put(651,176){\rule[-0.500pt]{2.168pt}{1.000pt}}
\put(660,175){\rule[-0.500pt]{3.373pt}{1.000pt}}
\put(674,174){\rule[-0.500pt]{3.373pt}{1.000pt}}
\put(688,173){\rule[-0.500pt]{3.132pt}{1.000pt}}
\put(701,172){\rule[-0.500pt]{3.373pt}{1.000pt}}
\end{picture}
\hspace{-.7cm} 
\setlength{\unitlength}{0.240900pt}
\ifx\plotpoint\undefined\newsavebox{\plotpoint}\fi
\sbox{\plotpoint}{\rule[-0.175pt]{0.350pt}{0.350pt}}%
\begin{picture}(779,600)(0,0)
\tenrm
\sbox{\plotpoint}{\rule[-0.175pt]{0.350pt}{0.350pt}}%
\put(264,290){\rule[-0.175pt]{4.818pt}{0.350pt}}
\put(242,290){\makebox(0,0)[r]{\small $20$}}
\put(695,290){\rule[-0.175pt]{4.818pt}{0.350pt}}
\put(264,421){\rule[-0.175pt]{4.818pt}{0.350pt}}
\put(242,421){\makebox(0,0)[r]{\small $40$}}
\put(695,421){\rule[-0.175pt]{4.818pt}{0.350pt}}
\put(397,158){\rule[-0.175pt]{0.350pt}{4.818pt}}
\put(397,113){\makebox(0,0){\small $5$}}
\put(397,467){\rule[-0.175pt]{0.350pt}{4.818pt}}
\put(529,158){\rule[-0.175pt]{0.350pt}{4.818pt}}
\put(529,113){\makebox(0,0){\small $10$}}
\put(529,467){\rule[-0.175pt]{0.350pt}{4.818pt}}
\put(662,158){\rule[-0.175pt]{0.350pt}{4.818pt}}
\put(662,113){\makebox(0,0){\small $15$}}
\put(662,467){\rule[-0.175pt]{0.350pt}{4.818pt}}
\put(264,158){\rule[-0.175pt]{108.646pt}{0.350pt}}
\put(715,158){\rule[-0.175pt]{0.350pt}{79.256pt}}
\put(264,487){\rule[-0.175pt]{108.646pt}{0.350pt}}
\put(67,322){\makebox(0,0)[l]{\shortstack{\large $E_{\rm o}^{v}$,$E_b$ \\ (meV)}}}
\put(533,68){\makebox(0,0){\Large $\alpha$ ($^{\rm o}$)}}
\put(635,336){\makebox(0,0)[l]{\large $E_{\rm o}^{v}$}}
\put(503,421){\makebox(0,0)[l]{\large $E_b$}}
\put(264,158){\rule[-0.175pt]{0.350pt}{79.256pt}}
\sbox{\plotpoint}{\rule[-0.500pt]{1.000pt}{1.000pt}}%
\put(264,158){\usebox{\plotpoint}}
\put(264,158){\rule[-0.500pt]{1.204pt}{1.000pt}}
\put(269,159){\usebox{\plotpoint}}
\put(271,160){\usebox{\plotpoint}}
\put(273,161){\rule[-0.500pt]{1.204pt}{1.000pt}}
\put(278,162){\usebox{\plotpoint}}
\put(280,163){\usebox{\plotpoint}}
\put(282,164){\rule[-0.500pt]{1.204pt}{1.000pt}}
\put(287,165){\usebox{\plotpoint}}
\put(291,166){\usebox{\plotpoint}}
\put(293,167){\usebox{\plotpoint}}
\put(296,168){\usebox{\plotpoint}}
\put(300,169){\rule[-0.500pt]{1.204pt}{1.000pt}}
\put(305,170){\usebox{\plotpoint}}
\put(307,171){\usebox{\plotpoint}}
\put(310,172){\usebox{\plotpoint}}
\put(314,173){\usebox{\plotpoint}}
\put(316,174){\usebox{\plotpoint}}
\put(319,175){\usebox{\plotpoint}}
\put(323,176){\rule[-0.500pt]{1.204pt}{1.000pt}}
\put(328,177){\usebox{\plotpoint}}
\put(330,178){\usebox{\plotpoint}}
\put(332,179){\rule[-0.500pt]{1.204pt}{1.000pt}}
\put(337,180){\usebox{\plotpoint}}
\put(341,181){\usebox{\plotpoint}}
\put(343,182){\usebox{\plotpoint}}
\put(346,183){\rule[-0.500pt]{1.204pt}{1.000pt}}
\put(351,184){\usebox{\plotpoint}}
\put(353,185){\usebox{\plotpoint}}
\put(355,186){\rule[-0.500pt]{1.204pt}{1.000pt}}
\put(360,187){\usebox{\plotpoint}}
\put(364,188){\usebox{\plotpoint}}
\put(366,189){\usebox{\plotpoint}}
\put(369,190){\usebox{\plotpoint}}
\put(373,191){\rule[-0.500pt]{1.204pt}{1.000pt}}
\put(378,192){\usebox{\plotpoint}}
\put(380,193){\usebox{\plotpoint}}
\put(382,194){\rule[-0.500pt]{1.204pt}{1.000pt}}
\put(387,195){\usebox{\plotpoint}}
\put(389,196){\usebox{\plotpoint}}
\put(392,197){\usebox{\plotpoint}}
\put(396,198){\rule[-0.500pt]{1.204pt}{1.000pt}}
\put(401,199){\usebox{\plotpoint}}
\put(403,200){\usebox{\plotpoint}}
\put(405,201){\rule[-0.500pt]{1.204pt}{1.000pt}}
\put(410,202){\usebox{\plotpoint}}
\put(414,203){\usebox{\plotpoint}}
\put(416,204){\usebox{\plotpoint}}
\put(419,205){\usebox{\plotpoint}}
\put(423,206){\usebox{\plotpoint}}
\put(425,207){\usebox{\plotpoint}}
\put(428,208){\rule[-0.500pt]{1.204pt}{1.000pt}}
\put(433,209){\usebox{\plotpoint}}
\put(437,210){\usebox{\plotpoint}}
\put(439,211){\usebox{\plotpoint}}
\put(442,212){\usebox{\plotpoint}}
\put(446,213){\rule[-0.500pt]{1.204pt}{1.000pt}}
\put(451,214){\usebox{\plotpoint}}
\put(453,215){\usebox{\plotpoint}}
\put(455,216){\rule[-0.500pt]{1.204pt}{1.000pt}}
\put(460,217){\usebox{\plotpoint}}
\put(462,218){\usebox{\plotpoint}}
\put(464,219){\rule[-0.500pt]{1.204pt}{1.000pt}}
\put(469,220){\rule[-0.500pt]{1.204pt}{1.000pt}}
\put(474,221){\usebox{\plotpoint}}
\put(476,222){\usebox{\plotpoint}}
\put(478,223){\rule[-0.500pt]{1.204pt}{1.000pt}}
\put(483,224){\usebox{\plotpoint}}
\put(487,225){\usebox{\plotpoint}}
\put(489,226){\usebox{\plotpoint}}
\put(492,227){\usebox{\plotpoint}}
\put(496,228){\usebox{\plotpoint}}
\put(498,229){\usebox{\plotpoint}}
\put(501,230){\usebox{\plotpoint}}
\put(505,231){\rule[-0.500pt]{1.204pt}{1.000pt}}
\put(510,232){\usebox{\plotpoint}}
\put(512,233){\usebox{\plotpoint}}
\put(515,234){\usebox{\plotpoint}}
\put(519,235){\rule[-0.500pt]{1.204pt}{1.000pt}}
\put(524,236){\usebox{\plotpoint}}
\put(526,237){\usebox{\plotpoint}}
\put(528,238){\rule[-0.500pt]{1.204pt}{1.000pt}}
\put(533,239){\usebox{\plotpoint}}
\put(537,240){\usebox{\plotpoint}}
\put(539,241){\usebox{\plotpoint}}
\put(542,242){\usebox{\plotpoint}}
\put(546,243){\rule[-0.500pt]{1.204pt}{1.000pt}}
\put(551,244){\usebox{\plotpoint}}
\put(553,245){\usebox{\plotpoint}}
\put(556,246){\usebox{\plotpoint}}
\put(560,247){\usebox{\plotpoint}}
\put(562,248){\usebox{\plotpoint}}
\put(565,249){\usebox{\plotpoint}}
\put(569,250){\rule[-0.500pt]{1.204pt}{1.000pt}}
\put(574,251){\usebox{\plotpoint}}
\put(576,252){\usebox{\plotpoint}}
\put(578,253){\rule[-0.500pt]{1.204pt}{1.000pt}}
\put(583,254){\usebox{\plotpoint}}
\put(587,255){\usebox{\plotpoint}}
\put(589,256){\usebox{\plotpoint}}
\put(592,257){\rule[-0.500pt]{1.204pt}{1.000pt}}
\put(597,258){\usebox{\plotpoint}}
\put(601,259){\usebox{\plotpoint}}
\put(603,260){\usebox{\plotpoint}}
\put(606,261){\usebox{\plotpoint}}
\put(610,262){\rule[-0.500pt]{1.204pt}{1.000pt}}
\put(615,263){\usebox{\plotpoint}}
\put(617,264){\usebox{\plotpoint}}
\put(619,265){\rule[-0.500pt]{1.204pt}{1.000pt}}
\put(624,266){\usebox{\plotpoint}}
\put(628,267){\usebox{\plotpoint}}
\put(630,268){\usebox{\plotpoint}}
\put(633,269){\rule[-0.500pt]{1.204pt}{1.000pt}}
\put(638,270){\usebox{\plotpoint}}
\put(642,271){\usebox{\plotpoint}}
\put(644,272){\usebox{\plotpoint}}
\put(647,273){\usebox{\plotpoint}}
\put(651,274){\rule[-0.500pt]{1.204pt}{1.000pt}}
\put(656,275){\usebox{\plotpoint}}
\put(658,276){\usebox{\plotpoint}}
\put(660,277){\rule[-0.500pt]{1.204pt}{1.000pt}}
\put(665,278){\usebox{\plotpoint}}
\put(669,279){\usebox{\plotpoint}}
\put(671,280){\usebox{\plotpoint}}
\put(674,281){\rule[-0.500pt]{1.204pt}{1.000pt}}
\put(679,282){\usebox{\plotpoint}}
\put(683,283){\usebox{\plotpoint}}
\put(685,284){\usebox{\plotpoint}}
\put(688,285){\usebox{\plotpoint}}
\put(692,286){\rule[-0.500pt]{1.204pt}{1.000pt}}
\put(697,287){\usebox{\plotpoint}}
\put(699,288){\usebox{\plotpoint}}
\put(701,289){\rule[-0.500pt]{1.204pt}{1.000pt}}
\put(706,290){\usebox{\plotpoint}}
\put(710,291){\usebox{\plotpoint}}
\put(712,292){\usebox{\plotpoint}}
\put(264,158){\usebox{\plotpoint}}
\put(264,158){\rule[-0.500pt]{16.381pt}{1.000pt}}
\put(332,159){\rule[-0.500pt]{5.541pt}{1.000pt}}
\put(355,160){\rule[-0.500pt]{3.373pt}{1.000pt}}
\put(369,161){\usebox{\plotpoint}}
\put(373,162){\rule[-0.500pt]{2.168pt}{1.000pt}}
\put(382,163){\rule[-0.500pt]{1.204pt}{1.000pt}}
\put(387,164){\rule[-0.500pt]{1.204pt}{1.000pt}}
\put(392,165){\usebox{\plotpoint}}
\put(396,166){\rule[-0.500pt]{1.204pt}{1.000pt}}
\put(401,167){\usebox{\plotpoint}}
\put(405,168){\rule[-0.500pt]{1.204pt}{1.000pt}}
\put(410,169){\usebox{\plotpoint}}
\put(412,170){\usebox{\plotpoint}}
\put(414,171){\rule[-0.500pt]{1.204pt}{1.000pt}}
\put(419,172){\usebox{\plotpoint}}
\put(421,173){\usebox{\plotpoint}}
\put(423,174){\usebox{\plotpoint}}
\put(425,175){\usebox{\plotpoint}}
\put(428,176){\usebox{\plotpoint}}
\put(430,177){\usebox{\plotpoint}}
\put(433,178){\usebox{\plotpoint}}
\put(435,179){\usebox{\plotpoint}}
\put(437,180){\usebox{\plotpoint}}
\put(438,181){\usebox{\plotpoint}}
\put(440,182){\usebox{\plotpoint}}
\put(441,183){\usebox{\plotpoint}}
\put(444,184){\usebox{\plotpoint}}
\put(446,185){\usebox{\plotpoint}}
\put(447,186){\usebox{\plotpoint}}
\put(449,187){\usebox{\plotpoint}}
\put(450,188){\usebox{\plotpoint}}
\put(452,189){\usebox{\plotpoint}}
\put(453,190){\usebox{\plotpoint}}
\put(455,191){\usebox{\plotpoint}}
\put(456,192){\usebox{\plotpoint}}
\put(458,193){\usebox{\plotpoint}}
\put(459,194){\usebox{\plotpoint}}
\put(461,195){\usebox{\plotpoint}}
\put(462,196){\usebox{\plotpoint}}
\put(463,197){\usebox{\plotpoint}}
\put(464,198){\usebox{\plotpoint}}
\put(465,199){\usebox{\plotpoint}}
\put(467,200){\usebox{\plotpoint}}
\put(468,201){\usebox{\plotpoint}}
\put(470,202){\usebox{\plotpoint}}
\put(471,203){\usebox{\plotpoint}}
\put(472,204){\usebox{\plotpoint}}
\put(474,205){\usebox{\plotpoint}}
\put(475,206){\usebox{\plotpoint}}
\put(476,207){\usebox{\plotpoint}}
\put(477,208){\usebox{\plotpoint}}
\put(478,210){\usebox{\plotpoint}}
\put(479,211){\usebox{\plotpoint}}
\put(480,212){\usebox{\plotpoint}}
\put(481,213){\usebox{\plotpoint}}
\put(483,214){\usebox{\plotpoint}}
\put(484,215){\usebox{\plotpoint}}
\put(485,216){\usebox{\plotpoint}}
\put(486,217){\usebox{\plotpoint}}
\put(487,219){\usebox{\plotpoint}}
\put(488,220){\usebox{\plotpoint}}
\put(489,221){\usebox{\plotpoint}}
\put(490,222){\usebox{\plotpoint}}
\put(491,223){\usebox{\plotpoint}}
\put(492,224){\usebox{\plotpoint}}
\put(493,225){\usebox{\plotpoint}}
\put(494,227){\usebox{\plotpoint}}
\put(495,228){\usebox{\plotpoint}}
\put(496,230){\usebox{\plotpoint}}
\put(497,231){\usebox{\plotpoint}}
\put(498,232){\usebox{\plotpoint}}
\put(499,233){\usebox{\plotpoint}}
\put(500,234){\usebox{\plotpoint}}
\put(501,235){\usebox{\plotpoint}}
\put(502,236){\usebox{\plotpoint}}
\put(503,238){\usebox{\plotpoint}}
\put(504,239){\usebox{\plotpoint}}
\put(505,241){\usebox{\plotpoint}}
\put(506,242){\usebox{\plotpoint}}
\put(507,243){\usebox{\plotpoint}}
\put(508,245){\usebox{\plotpoint}}
\put(509,246){\usebox{\plotpoint}}
\put(510,247){\usebox{\plotpoint}}
\put(511,249){\usebox{\plotpoint}}
\put(512,250){\usebox{\plotpoint}}
\put(513,252){\usebox{\plotpoint}}
\put(514,253){\usebox{\plotpoint}}
\put(515,254){\usebox{\plotpoint}}
\put(516,256){\usebox{\plotpoint}}
\put(517,258){\usebox{\plotpoint}}
\put(518,260){\usebox{\plotpoint}}
\put(519,262){\usebox{\plotpoint}}
\put(520,263){\usebox{\plotpoint}}
\put(521,265){\usebox{\plotpoint}}
\put(522,266){\usebox{\plotpoint}}
\put(523,268){\usebox{\plotpoint}}
\put(524,270){\usebox{\plotpoint}}
\put(525,272){\usebox{\plotpoint}}
\put(526,274){\usebox{\plotpoint}}
\put(527,276){\usebox{\plotpoint}}
\put(528,278){\usebox{\plotpoint}}
\put(529,279){\usebox{\plotpoint}}
\put(530,281){\usebox{\plotpoint}}
\put(531,282){\usebox{\plotpoint}}
\put(532,284){\usebox{\plotpoint}}
\put(533,286){\usebox{\plotpoint}}
\put(534,288){\usebox{\plotpoint}}
\put(535,290){\usebox{\plotpoint}}
\put(536,292){\usebox{\plotpoint}}
\put(537,295){\usebox{\plotpoint}}
\put(538,296){\usebox{\plotpoint}}
\put(539,298){\usebox{\plotpoint}}
\put(540,300){\usebox{\plotpoint}}
\put(541,302){\usebox{\plotpoint}}
\put(542,303){\usebox{\plotpoint}}
\put(543,306){\usebox{\plotpoint}}
\put(544,309){\usebox{\plotpoint}}
\put(545,311){\usebox{\plotpoint}}
\put(546,314){\usebox{\plotpoint}}
\put(547,316){\usebox{\plotpoint}}
\put(548,318){\usebox{\plotpoint}}
\put(549,320){\usebox{\plotpoint}}
\put(550,322){\usebox{\plotpoint}}
\put(551,324){\usebox{\plotpoint}}
\put(552,326){\usebox{\plotpoint}}
\put(553,328){\usebox{\plotpoint}}
\put(554,330){\usebox{\plotpoint}}
\put(555,332){\usebox{\plotpoint}}
\put(556,335){\usebox{\plotpoint}}
\put(557,337){\usebox{\plotpoint}}
\put(558,340){\usebox{\plotpoint}}
\put(559,343){\usebox{\plotpoint}}
\put(560,346){\usebox{\plotpoint}}
\put(561,348){\usebox{\plotpoint}}
\put(562,350){\usebox{\plotpoint}}
\put(563,353){\usebox{\plotpoint}}
\put(564,355){\usebox{\plotpoint}}
\put(565,357){\usebox{\plotpoint}}
\put(566,361){\usebox{\plotpoint}}
\put(567,364){\usebox{\plotpoint}}
\put(568,367){\usebox{\plotpoint}}
\put(569,371){\usebox{\plotpoint}}
\put(570,373){\usebox{\plotpoint}}
\put(571,376){\usebox{\plotpoint}}
\put(572,378){\usebox{\plotpoint}}
\put(573,381){\usebox{\plotpoint}}
\put(574,384){\usebox{\plotpoint}}
\put(575,387){\usebox{\plotpoint}}
\put(576,390){\usebox{\plotpoint}}
\put(577,393){\usebox{\plotpoint}}
\put(578,397){\usebox{\plotpoint}}
\put(579,399){\usebox{\plotpoint}}
\put(580,402){\usebox{\plotpoint}}
\put(581,405){\usebox{\plotpoint}}
\put(582,408){\usebox{\plotpoint}}
\put(583,410){\usebox{\plotpoint}}
\put(584,414){\usebox{\plotpoint}}
\put(585,418){\usebox{\plotpoint}}
\put(586,422){\usebox{\plotpoint}}
\put(587,426){\usebox{\plotpoint}}
\put(588,429){\usebox{\plotpoint}}
\put(589,432){\usebox{\plotpoint}}
\put(590,435){\usebox{\plotpoint}}
\put(591,438){\usebox{\plotpoint}}
\put(592,441){\usebox{\plotpoint}}
\put(593,444){\usebox{\plotpoint}}
\put(594,447){\usebox{\plotpoint}}
\put(595,450){\usebox{\plotpoint}}
\put(596,453){\usebox{\plotpoint}}
\put(597,457){\rule[-0.500pt]{1.000pt}{1.024pt}}
\put(598,461){\rule[-0.500pt]{1.000pt}{1.024pt}}
\put(599,465){\rule[-0.500pt]{1.000pt}{1.024pt}}
\put(600,469){\rule[-0.500pt]{1.000pt}{1.024pt}}
\put(601,474){\usebox{\plotpoint}}
\put(602,477){\usebox{\plotpoint}}
\put(603,480){\usebox{\plotpoint}}
\put(604,483){\usebox{\plotpoint}}
\end{picture}
} 
\vspace{-1.3cm}
\end{center}
\caption{(a) Dependence of $t$ and $B$ on the puckering angle $\alpha$,
(b) The energy barrier $E_b$ and lowest vibrational
energy $E_{\rm o}^v$ vs. $\alpha$ for the radial
motion of oxygen. SiO$_2$ case.}
\label{fig:par}
\vspace{-15pt}
\end{figure}
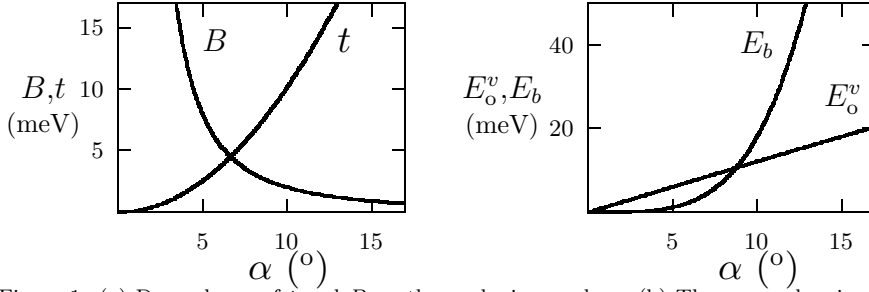

\vspace{-4pt}

\section{Model and Hamiltonian}

\vspace{-3pt}

A lattice of fixed $X$ atoms is considered with O atoms 
bridging between them. The degrees of 
freedom are the angles $\phi_i$ associated with the oxygens in their 
rotations around the axes defined by the neighboring nodes (given a fixed
$X$-O distance those rotations are the only possible motions). The 
Hamiltonian is 
\begin{equation}
\vspace{-10pt}
{\cal H}=-B \sum_i \partial^2 / \partial \phi_i^2 +2t \sum_{\langle i,j \rangle}
\cos(\phi_i-\phi_j) -2 C\sum_i \cos(n \phi_i),
\label{ham}
\vspace{-2pt}
\end{equation}
including the kinetic energy, a nearest 
neighbor interaction which tends to align the O bridges ($t>0$), 
and a hindering potential. The latter accounts for effects in 
real solids which impede the rotation of the oxygen atoms. It
is a fixed potential, whereas in the real solids the impeding
objects (other atoms) are mobile. $n$ gives the number of potential minima
around the circle. 

    The rotor constant is $B\! =\! \hbar^2 / 2I$, where $I\! =\! m_{\rm O} 
R^2$, and $R\! =\! b \sin \alpha$, $b$ being the bond length (1.6 \AA\ 
for Si-O, 1.9 for Cu-O), and the puckering angle 
$\alpha\! =\! (\pi\!\! -\!\! 2 \, \theta_{\mbox{\small {\em X}-O-{\em X}}})/2$. 
Therefore, $B\! =\! B_{\rm o}/\sin^{2} \alpha$, with $B_{\rm o}\!\approx\! 
0.06$ meV for Si and 0.04 for Cu.
The interaction parameter $t$ accounts for the energy needed to 
rotate an O atom while keeping its neighbors fixed. This energy is essentially 
used to bend the O-$X$-O angle away from its natural value. Then,
$t=t_{\rm o} (1-\cos 2\alpha)$. Using the O-$X$-O bond bending force 
constants, $t_{\rm o} \approx 170$
meV for Si, and 230 for Cu. $B$ decreases with $\alpha$ whereas
$t$ increases, defining an $\alpha$ for which $B=t$ (7$^{\rm o}$ for Si,
5$^{\rm o}$ for Cu). $C$ is much more dependent on 
particular crystal structures and more difficult to estimate. It 
increases with $\alpha$, and can be as small as 0.1 meV 
for $\alpha=20^{\rm o}$, as found for interstitial O in Ge.\cite{japon}

    The radial vibration of the O atoms involves the stretching of 
$X$-O bonds and can be neglected for well-defined rotor situations.
However, for small enough $\alpha$, the radial motion
becomes important since $\Delta R \sim \langle R \rangle$. 
An estimate of the range of validity of ${\cal H}$ can be drawn from the
$X$-O bond stretching force constants, by comparing the ground-state
energy for the radial vibrations $E_{\rm o}^v$ with $E_b$, the potential energy 
barrier at $R=0$. $E_b= K (1-\cos\alpha)^2$, with $K \approx 80$ eV
for Si and 30 for Cu. $E_{\rm o}^v = K^{\prime} \sin\alpha$, with
$K^{\prime} \approx 70$ meV for Si, 35 for Cu. 
$E_b > E_{\rm o}^v$ for $\alpha > 9^{\rm o}$ both for Si and Cu.
Only the $\phi_i$ will be considered here, the 
effects of the radial motion being the subject of future work.\cite{we}

\begin{figure}
\begin{center}
\mbox{ \hspace {-2.cm} 
\setlength{\unitlength}{0.240900pt}
\ifx\plotpoint\undefined\newsavebox{\plotpoint}\fi
\sbox{\plotpoint}{\rule[-0.175pt]{0.350pt}{0.350pt}}%
\begin{picture}(719,660)(0,0)
\tenrm
\sbox{\plotpoint}{\rule[-0.175pt]{0.350pt}{0.350pt}}%
\put(264,158){\rule[-0.175pt]{94.192pt}{0.350pt}}
\put(655,158){\rule[-0.175pt]{0.350pt}{93.710pt}}
\put(264,547){\rule[-0.175pt]{94.192pt}{0.350pt}}
\put(199,352){\makebox(0,0)[l]{\shortstack{\LARGE $\varepsilon$}}}
\put(459,113){\makebox(0,0){\large $k$}}
\put(467,244){\makebox(0,0)[l]{$\Delta \!\!\sim \!B\!-\!2t$}}
\put(287,298){\makebox(0,0)[l]{$W \sim 4t$}}
\put(303,504){\makebox(0,0)[l]{\Large (a) $t$ \large $\ll B$}}
\put(604,130){\makebox(0,0)[l]{$\pi / a$}}
\put(244,130){\makebox(0,0)[l]{$-\pi / a$}}
\put(225,167){\makebox(0,0)[l]{\small 0}}
\put(264,158){\rule[-0.175pt]{0.350pt}{93.710pt}}
\put(460,158){\vector(0,1){167}}
\put(460,325){\vector(0,-1){167}}
\put(284,325){\vector(0,1){89}}
\put(284,414){\vector(0,-1){89}}
\put(272,325){\line(1,0){39}}
\sbox{\plotpoint}{\rule[-0.500pt]{1.000pt}{1.000pt}}%
\put(264,414){\usebox{\plotpoint}}
\put(264,414){\rule[-0.500pt]{1.927pt}{1.000pt}}
\put(272,413){\rule[-0.500pt]{1.927pt}{1.000pt}}
\put(280,412){\rule[-0.500pt]{1.927pt}{1.000pt}}
\put(288,411){\usebox{\plotpoint}}
\put(292,410){\usebox{\plotpoint}}
\put(296,409){\usebox{\plotpoint}}
\put(300,408){\usebox{\plotpoint}}
\put(301,407){\usebox{\plotpoint}}
\put(303,406){\usebox{\plotpoint}}
\put(307,405){\usebox{\plotpoint}}
\put(309,404){\usebox{\plotpoint}}
\put(311,403){\usebox{\plotpoint}}
\put(313,402){\usebox{\plotpoint}}
\put(315,401){\usebox{\plotpoint}}
\put(317,400){\usebox{\plotpoint}}
\put(319,399){\usebox{\plotpoint}}
\put(321,398){\usebox{\plotpoint}}
\put(323,397){\usebox{\plotpoint}}
\put(325,396){\usebox{\plotpoint}}
\put(327,395){\usebox{\plotpoint}}
\put(329,394){\usebox{\plotpoint}}
\put(331,393){\usebox{\plotpoint}}
\put(333,392){\usebox{\plotpoint}}
\put(335,391){\usebox{\plotpoint}}
\put(337,390){\usebox{\plotpoint}}
\put(339,389){\usebox{\plotpoint}}
\put(340,388){\usebox{\plotpoint}}
\put(341,387){\usebox{\plotpoint}}
\put(343,386){\usebox{\plotpoint}}
\put(345,385){\usebox{\plotpoint}}
\put(347,384){\usebox{\plotpoint}}
\put(348,383){\usebox{\plotpoint}}
\put(349,382){\usebox{\plotpoint}}
\put(351,381){\usebox{\plotpoint}}
\put(352,380){\usebox{\plotpoint}}
\put(353,379){\usebox{\plotpoint}}
\put(355,378){\usebox{\plotpoint}}
\put(357,377){\usebox{\plotpoint}}
\put(359,376){\usebox{\plotpoint}}
\put(360,375){\usebox{\plotpoint}}
\put(361,374){\usebox{\plotpoint}}
\put(363,373){\usebox{\plotpoint}}
\put(364,372){\usebox{\plotpoint}}
\put(365,371){\usebox{\plotpoint}}
\put(367,370){\usebox{\plotpoint}}
\put(368,369){\usebox{\plotpoint}}
\put(369,368){\usebox{\plotpoint}}
\put(371,367){\usebox{\plotpoint}}
\put(373,366){\usebox{\plotpoint}}
\put(375,365){\usebox{\plotpoint}}
\put(376,364){\usebox{\plotpoint}}
\put(377,363){\usebox{\plotpoint}}
\put(379,362){\usebox{\plotpoint}}
\put(380,361){\usebox{\plotpoint}}
\put(381,360){\usebox{\plotpoint}}
\put(382,359){\usebox{\plotpoint}}
\put(383,358){\usebox{\plotpoint}}
\put(384,357){\usebox{\plotpoint}}
\put(386,356){\usebox{\plotpoint}}
\put(388,355){\usebox{\plotpoint}}
\put(390,354){\usebox{\plotpoint}}
\put(391,353){\usebox{\plotpoint}}
\put(392,352){\usebox{\plotpoint}}
\put(394,351){\usebox{\plotpoint}}
\put(395,350){\usebox{\plotpoint}}
\put(396,349){\usebox{\plotpoint}}
\put(398,348){\usebox{\plotpoint}}
\put(400,347){\usebox{\plotpoint}}
\put(402,346){\usebox{\plotpoint}}
\put(403,345){\usebox{\plotpoint}}
\put(404,344){\usebox{\plotpoint}}
\put(406,343){\usebox{\plotpoint}}
\put(408,342){\usebox{\plotpoint}}
\put(410,341){\usebox{\plotpoint}}
\put(412,340){\usebox{\plotpoint}}
\put(414,339){\usebox{\plotpoint}}
\put(416,338){\usebox{\plotpoint}}
\put(418,337){\usebox{\plotpoint}}
\put(420,336){\usebox{\plotpoint}}
\put(422,335){\usebox{\plotpoint}}
\put(424,334){\usebox{\plotpoint}}
\put(426,333){\usebox{\plotpoint}}
\put(428,332){\usebox{\plotpoint}}
\put(430,331){\usebox{\plotpoint}}
\put(434,330){\usebox{\plotpoint}}
\put(438,329){\usebox{\plotpoint}}
\put(442,328){\usebox{\plotpoint}}
\put(446,327){\usebox{\plotpoint}}
\put(450,326){\rule[-0.500pt]{1.927pt}{1.000pt}}
\put(458,325){\rule[-0.500pt]{1.686pt}{1.000pt}}
\put(465,326){\rule[-0.500pt]{1.927pt}{1.000pt}}
\put(473,327){\usebox{\plotpoint}}
\put(477,328){\usebox{\plotpoint}}
\put(481,329){\usebox{\plotpoint}}
\put(485,330){\usebox{\plotpoint}}
\put(489,331){\usebox{\plotpoint}}
\put(491,332){\usebox{\plotpoint}}
\put(493,333){\usebox{\plotpoint}}
\put(495,334){\usebox{\plotpoint}}
\put(497,335){\usebox{\plotpoint}}
\put(499,336){\usebox{\plotpoint}}
\put(501,337){\usebox{\plotpoint}}
\put(503,338){\usebox{\plotpoint}}
\put(505,339){\usebox{\plotpoint}}
\put(507,340){\usebox{\plotpoint}}
\put(509,341){\usebox{\plotpoint}}
\put(511,342){\usebox{\plotpoint}}
\put(513,343){\usebox{\plotpoint}}
\put(514,344){\usebox{\plotpoint}}
\put(515,345){\usebox{\plotpoint}}
\put(516,346){\usebox{\plotpoint}}
\put(519,347){\usebox{\plotpoint}}
\put(521,348){\usebox{\plotpoint}}
\put(522,349){\usebox{\plotpoint}}
\put(523,350){\usebox{\plotpoint}}
\put(524,351){\usebox{\plotpoint}}
\put(526,352){\usebox{\plotpoint}}
\put(527,353){\usebox{\plotpoint}}
\put(528,354){\usebox{\plotpoint}}
\put(531,355){\usebox{\plotpoint}}
\put(533,356){\usebox{\plotpoint}}
\put(534,357){\usebox{\plotpoint}}
\put(535,358){\usebox{\plotpoint}}
\put(536,359){\usebox{\plotpoint}}
\put(538,360){\usebox{\plotpoint}}
\put(539,361){\usebox{\plotpoint}}
\put(540,362){\usebox{\plotpoint}}
\put(541,363){\usebox{\plotpoint}}
\put(542,364){\usebox{\plotpoint}}
\put(543,365){\usebox{\plotpoint}}
\put(546,366){\usebox{\plotpoint}}
\put(548,367){\usebox{\plotpoint}}
\put(549,368){\usebox{\plotpoint}}
\put(550,369){\usebox{\plotpoint}}
\put(551,370){\usebox{\plotpoint}}
\put(553,371){\usebox{\plotpoint}}
\put(554,372){\usebox{\plotpoint}}
\put(555,373){\usebox{\plotpoint}}
\put(557,374){\usebox{\plotpoint}}
\put(558,375){\usebox{\plotpoint}}
\put(559,376){\usebox{\plotpoint}}
\put(562,377){\usebox{\plotpoint}}
\put(564,378){\usebox{\plotpoint}}
\put(565,379){\usebox{\plotpoint}}
\put(566,380){\usebox{\plotpoint}}
\put(567,381){\usebox{\plotpoint}}
\put(569,382){\usebox{\plotpoint}}
\put(570,383){\usebox{\plotpoint}}
\put(571,384){\usebox{\plotpoint}}
\put(574,385){\usebox{\plotpoint}}
\put(576,386){\usebox{\plotpoint}}
\put(577,387){\usebox{\plotpoint}}
\put(578,388){\usebox{\plotpoint}}
\put(579,389){\usebox{\plotpoint}}
\put(582,390){\usebox{\plotpoint}}
\put(584,391){\usebox{\plotpoint}}
\put(586,392){\usebox{\plotpoint}}
\put(588,393){\usebox{\plotpoint}}
\put(590,394){\usebox{\plotpoint}}
\put(592,395){\usebox{\plotpoint}}
\put(594,396){\usebox{\plotpoint}}
\put(596,397){\usebox{\plotpoint}}
\put(598,398){\usebox{\plotpoint}}
\put(600,399){\usebox{\plotpoint}}
\put(602,400){\usebox{\plotpoint}}
\put(604,401){\usebox{\plotpoint}}
\put(606,402){\usebox{\plotpoint}}
\put(608,403){\usebox{\plotpoint}}
\put(610,404){\usebox{\plotpoint}}
\put(612,405){\usebox{\plotpoint}}
\put(616,406){\usebox{\plotpoint}}
\put(617,407){\usebox{\plotpoint}}
\put(619,408){\usebox{\plotpoint}}
\put(623,409){\usebox{\plotpoint}}
\put(627,410){\usebox{\plotpoint}}
\put(631,411){\usebox{\plotpoint}}
\put(635,412){\rule[-0.500pt]{1.927pt}{1.000pt}}
\put(643,413){\rule[-0.500pt]{1.927pt}{1.000pt}}
\put(651,414){\usebox{\plotpoint}}
\end{picture}
\hspace{-2.8cm} 
\setlength{\unitlength}{0.240900pt}
\ifx\plotpoint\undefined\newsavebox{\plotpoint}\fi
\sbox{\plotpoint}{\rule[-0.175pt]{0.350pt}{0.350pt}}%
\begin{picture}(719,660)(0,0)
\tenrm
\sbox{\plotpoint}{\rule[-0.175pt]{0.350pt}{0.350pt}}%
\put(264,158){\rule[-0.175pt]{94.192pt}{0.350pt}}
\put(655,158){\rule[-0.175pt]{0.350pt}{93.710pt}}
\put(264,547){\rule[-0.175pt]{94.192pt}{0.350pt}}
\put(459,113){\makebox(0,0){\large $k$}}
\put(280,190){\makebox(0,0)[l]{\small $W \!\! \! \sim \!\! 4 \! \sqrt {\! B t}$}}
\put(624,130){\makebox(0,0)[l]{$\pi / a$}}
\put(256,130){\makebox(0,0)[l]{$-\pi / a$}}
\put(362,363){\makebox(0,0)[l]{{\large $v_s$}{\small $\sim\! 2\sqrt {Bt}$}{\large ${a\over\hbar}$}}}
\put(303,504){\makebox(0,0)[l]{\Large (b) $t$ \large $\gg B$}}
\put(264,158){\rule[-0.175pt]{0.350pt}{93.710pt}}
\put(276,158){\vector(0,1){216}}
\put(276,374){\vector(0,-1){216}}
\put(479,337){\rule[-0.175pt]{0.350pt}{1.180pt}}
\put(480,332){\rule[-0.175pt]{0.350pt}{1.180pt}}
\put(481,327){\rule[-0.175pt]{0.350pt}{1.180pt}}
\put(482,322){\rule[-0.175pt]{0.350pt}{1.180pt}}
\put(483,317){\rule[-0.175pt]{0.350pt}{1.180pt}}
\put(484,312){\rule[-0.175pt]{0.350pt}{1.180pt}}
\put(485,307){\rule[-0.175pt]{0.350pt}{1.180pt}}
\put(486,302){\rule[-0.175pt]{0.350pt}{1.180pt}}
\put(487,297){\rule[-0.175pt]{0.350pt}{1.180pt}}
\put(488,293){\rule[-0.175pt]{0.350pt}{1.180pt}}
\put(489,288){\rule[-0.175pt]{0.350pt}{1.180pt}}
\put(490,283){\rule[-0.175pt]{0.350pt}{1.180pt}}
\put(491,278){\rule[-0.175pt]{0.350pt}{1.180pt}}
\put(492,273){\rule[-0.175pt]{0.350pt}{1.180pt}}
\put(493,268){\rule[-0.175pt]{0.350pt}{1.180pt}}
\put(494,263){\rule[-0.175pt]{0.350pt}{1.180pt}}
\put(495,258){\rule[-0.175pt]{0.350pt}{1.180pt}}
\put(496,253){\rule[-0.175pt]{0.350pt}{1.180pt}}
\put(497,248){\rule[-0.175pt]{0.350pt}{1.180pt}}
\put(498,244){\rule[-0.175pt]{0.350pt}{1.180pt}}
\put(499,244){\vector(1,-4){0}}
\put(499,244){\usebox{\plotpoint}}
\sbox{\plotpoint}{\rule[-0.500pt]{1.000pt}{1.000pt}}%
\put(264,374){\usebox{\plotpoint}}
\put(264,374){\rule[-0.500pt]{2.891pt}{1.000pt}}
\put(276,373){\usebox{\plotpoint}}
\put(280,372){\usebox{\plotpoint}}
\put(284,371){\usebox{\plotpoint}}
\put(288,370){\usebox{\plotpoint}}
\put(292,369){\usebox{\plotpoint}}
\put(294,368){\usebox{\plotpoint}}
\put(296,367){\usebox{\plotpoint}}
\put(298,366){\usebox{\plotpoint}}
\put(300,365){\usebox{\plotpoint}}
\put(301,364){\usebox{\plotpoint}}
\put(303,363){\usebox{\plotpoint}}
\put(305,362){\usebox{\plotpoint}}
\put(307,361){\usebox{\plotpoint}}
\put(309,360){\usebox{\plotpoint}}
\put(311,359){\usebox{\plotpoint}}
\put(312,358){\usebox{\plotpoint}}
\put(313,357){\usebox{\plotpoint}}
\put(315,356){\usebox{\plotpoint}}
\put(316,355){\usebox{\plotpoint}}
\put(317,354){\usebox{\plotpoint}}
\put(319,353){\usebox{\plotpoint}}
\put(320,352){\usebox{\plotpoint}}
\put(321,351){\usebox{\plotpoint}}
\put(323,350){\usebox{\plotpoint}}
\put(324,349){\usebox{\plotpoint}}
\put(325,348){\usebox{\plotpoint}}
\put(327,347){\usebox{\plotpoint}}
\put(328,346){\usebox{\plotpoint}}
\put(329,345){\usebox{\plotpoint}}
\put(330,344){\usebox{\plotpoint}}
\put(331,343){\usebox{\plotpoint}}
\put(332,342){\usebox{\plotpoint}}
\put(333,341){\usebox{\plotpoint}}
\put(335,340){\usebox{\plotpoint}}
\put(336,339){\usebox{\plotpoint}}
\put(337,338){\usebox{\plotpoint}}
\put(338,337){\usebox{\plotpoint}}
\put(339,336){\usebox{\plotpoint}}
\put(340,335){\usebox{\plotpoint}}
\put(341,334){\usebox{\plotpoint}}
\put(342,333){\usebox{\plotpoint}}
\put(343,332){\usebox{\plotpoint}}
\put(344,331){\usebox{\plotpoint}}
\put(345,330){\usebox{\plotpoint}}
\put(346,329){\usebox{\plotpoint}}
\put(347,328){\usebox{\plotpoint}}
\put(348,327){\usebox{\plotpoint}}
\put(349,326){\usebox{\plotpoint}}
\put(350,325){\usebox{\plotpoint}}
\put(351,322){\usebox{\plotpoint}}
\put(352,321){\usebox{\plotpoint}}
\put(353,320){\usebox{\plotpoint}}
\put(354,319){\usebox{\plotpoint}}
\put(355,317){\usebox{\plotpoint}}
\put(356,316){\usebox{\plotpoint}}
\put(357,315){\usebox{\plotpoint}}
\put(358,314){\usebox{\plotpoint}}
\put(359,314){\usebox{\plotpoint}}
\put(360,313){\usebox{\plotpoint}}
\put(361,312){\usebox{\plotpoint}}
\put(362,311){\usebox{\plotpoint}}
\put(363,308){\usebox{\plotpoint}}
\put(364,307){\usebox{\plotpoint}}
\put(365,306){\usebox{\plotpoint}}
\put(366,305){\usebox{\plotpoint}}
\put(367,303){\usebox{\plotpoint}}
\put(368,302){\usebox{\plotpoint}}
\put(369,301){\usebox{\plotpoint}}
\put(370,300){\usebox{\plotpoint}}
\put(371,298){\usebox{\plotpoint}}
\put(372,297){\usebox{\plotpoint}}
\put(373,295){\usebox{\plotpoint}}
\put(374,294){\usebox{\plotpoint}}
\put(375,292){\usebox{\plotpoint}}
\put(376,291){\usebox{\plotpoint}}
\put(377,290){\usebox{\plotpoint}}
\put(378,289){\usebox{\plotpoint}}
\put(379,287){\usebox{\plotpoint}}
\put(380,285){\usebox{\plotpoint}}
\put(381,283){\usebox{\plotpoint}}
\put(382,281){\usebox{\plotpoint}}
\put(383,280){\usebox{\plotpoint}}
\put(384,279){\usebox{\plotpoint}}
\put(385,278){\usebox{\plotpoint}}
\put(386,276){\usebox{\plotpoint}}
\put(387,275){\usebox{\plotpoint}}
\put(388,273){\usebox{\plotpoint}}
\put(389,272){\usebox{\plotpoint}}
\put(390,270){\usebox{\plotpoint}}
\put(391,269){\usebox{\plotpoint}}
\put(392,267){\usebox{\plotpoint}}
\put(393,266){\usebox{\plotpoint}}
\put(394,264){\usebox{\plotpoint}}
\put(395,263){\usebox{\plotpoint}}
\put(396,261){\usebox{\plotpoint}}
\put(397,260){\usebox{\plotpoint}}
\put(398,258){\usebox{\plotpoint}}
\put(399,257){\usebox{\plotpoint}}
\put(400,255){\usebox{\plotpoint}}
\put(401,254){\usebox{\plotpoint}}
\put(402,252){\usebox{\plotpoint}}
\put(403,251){\usebox{\plotpoint}}
\put(404,249){\usebox{\plotpoint}}
\put(405,248){\usebox{\plotpoint}}
\put(406,246){\usebox{\plotpoint}}
\put(407,244){\usebox{\plotpoint}}
\put(408,242){\usebox{\plotpoint}}
\put(409,241){\usebox{\plotpoint}}
\put(410,239){\usebox{\plotpoint}}
\put(411,238){\usebox{\plotpoint}}
\put(412,236){\usebox{\plotpoint}}
\put(413,235){\usebox{\plotpoint}}
\put(414,233){\usebox{\plotpoint}}
\put(415,232){\usebox{\plotpoint}}
\put(416,230){\usebox{\plotpoint}}
\put(417,229){\usebox{\plotpoint}}
\put(418,227){\usebox{\plotpoint}}
\put(419,225){\usebox{\plotpoint}}
\put(420,223){\usebox{\plotpoint}}
\put(421,222){\usebox{\plotpoint}}
\put(422,220){\usebox{\plotpoint}}
\put(423,219){\usebox{\plotpoint}}
\put(424,217){\usebox{\plotpoint}}
\put(425,216){\usebox{\plotpoint}}
\put(426,214){\usebox{\plotpoint}}
\put(427,212){\usebox{\plotpoint}}
\put(428,210){\usebox{\plotpoint}}
\put(429,209){\usebox{\plotpoint}}
\put(430,207){\usebox{\plotpoint}}
\put(431,205){\usebox{\plotpoint}}
\put(432,203){\usebox{\plotpoint}}
\put(433,202){\usebox{\plotpoint}}
\put(434,200){\usebox{\plotpoint}}
\put(435,199){\usebox{\plotpoint}}
\put(436,197){\usebox{\plotpoint}}
\put(437,196){\usebox{\plotpoint}}
\put(438,194){\usebox{\plotpoint}}
\put(439,192){\usebox{\plotpoint}}
\put(440,190){\usebox{\plotpoint}}
\put(441,189){\usebox{\plotpoint}}
\put(442,187){\usebox{\plotpoint}}
\put(443,185){\usebox{\plotpoint}}
\put(444,183){\usebox{\plotpoint}}
\put(445,182){\usebox{\plotpoint}}
\put(446,180){\usebox{\plotpoint}}
\put(447,178){\usebox{\plotpoint}}
\put(448,176){\usebox{\plotpoint}}
\put(449,175){\usebox{\plotpoint}}
\put(450,173){\usebox{\plotpoint}}
\put(451,171){\usebox{\plotpoint}}
\put(452,169){\usebox{\plotpoint}}
\put(453,168){\usebox{\plotpoint}}
\put(454,166){\usebox{\plotpoint}}
\put(455,164){\usebox{\plotpoint}}
\put(456,162){\usebox{\plotpoint}}
\put(457,161){\usebox{\plotpoint}}
\put(458,161){\usebox{\plotpoint}}
\put(461,161){\usebox{\plotpoint}}
\put(462,162){\usebox{\plotpoint}}
\put(463,164){\usebox{\plotpoint}}
\put(464,166){\usebox{\plotpoint}}
\put(465,168){\usebox{\plotpoint}}
\put(466,169){\usebox{\plotpoint}}
\put(467,171){\usebox{\plotpoint}}
\put(468,173){\usebox{\plotpoint}}
\put(469,175){\usebox{\plotpoint}}
\put(470,176){\usebox{\plotpoint}}
\put(471,178){\usebox{\plotpoint}}
\put(472,180){\usebox{\plotpoint}}
\put(473,182){\usebox{\plotpoint}}
\put(474,183){\usebox{\plotpoint}}
\put(475,185){\usebox{\plotpoint}}
\put(476,187){\usebox{\plotpoint}}
\put(477,189){\usebox{\plotpoint}}
\put(478,190){\usebox{\plotpoint}}
\put(479,192){\usebox{\plotpoint}}
\put(480,194){\usebox{\plotpoint}}
\put(481,196){\usebox{\plotpoint}}
\put(482,197){\usebox{\plotpoint}}
\put(483,199){\usebox{\plotpoint}}
\put(484,200){\usebox{\plotpoint}}
\put(485,202){\usebox{\plotpoint}}
\put(486,203){\usebox{\plotpoint}}
\put(487,205){\usebox{\plotpoint}}
\put(488,207){\usebox{\plotpoint}}
\put(489,209){\usebox{\plotpoint}}
\put(490,210){\usebox{\plotpoint}}
\put(491,212){\usebox{\plotpoint}}
\put(492,214){\usebox{\plotpoint}}
\put(493,216){\usebox{\plotpoint}}
\put(494,217){\usebox{\plotpoint}}
\put(495,219){\usebox{\plotpoint}}
\put(496,220){\usebox{\plotpoint}}
\put(497,222){\usebox{\plotpoint}}
\put(498,223){\usebox{\plotpoint}}
\put(499,225){\usebox{\plotpoint}}
\put(500,227){\usebox{\plotpoint}}
\put(501,229){\usebox{\plotpoint}}
\put(502,230){\usebox{\plotpoint}}
\put(503,232){\usebox{\plotpoint}}
\put(504,233){\usebox{\plotpoint}}
\put(505,235){\usebox{\plotpoint}}
\put(506,236){\usebox{\plotpoint}}
\put(507,238){\usebox{\plotpoint}}
\put(508,239){\usebox{\plotpoint}}
\put(509,241){\usebox{\plotpoint}}
\put(510,242){\usebox{\plotpoint}}
\put(511,244){\usebox{\plotpoint}}
\put(512,246){\usebox{\plotpoint}}
\put(513,248){\usebox{\plotpoint}}
\put(514,249){\usebox{\plotpoint}}
\put(515,251){\usebox{\plotpoint}}
\put(516,252){\usebox{\plotpoint}}
\put(517,254){\usebox{\plotpoint}}
\put(518,255){\usebox{\plotpoint}}
\put(519,257){\usebox{\plotpoint}}
\put(520,258){\usebox{\plotpoint}}
\put(521,260){\usebox{\plotpoint}}
\put(522,261){\usebox{\plotpoint}}
\put(523,263){\usebox{\plotpoint}}
\put(524,264){\usebox{\plotpoint}}
\put(525,266){\usebox{\plotpoint}}
\put(526,267){\usebox{\plotpoint}}
\put(527,269){\usebox{\plotpoint}}
\put(528,270){\usebox{\plotpoint}}
\put(529,272){\usebox{\plotpoint}}
\put(530,273){\usebox{\plotpoint}}
\put(531,275){\usebox{\plotpoint}}
\put(532,276){\usebox{\plotpoint}}
\put(533,278){\usebox{\plotpoint}}
\put(534,279){\usebox{\plotpoint}}
\put(535,280){\usebox{\plotpoint}}
\put(536,281){\usebox{\plotpoint}}
\put(537,283){\usebox{\plotpoint}}
\put(538,285){\usebox{\plotpoint}}
\put(539,287){\usebox{\plotpoint}}
\put(540,289){\usebox{\plotpoint}}
\put(541,290){\usebox{\plotpoint}}
\put(542,291){\usebox{\plotpoint}}
\put(543,292){\usebox{\plotpoint}}
\put(544,294){\usebox{\plotpoint}}
\put(545,295){\usebox{\plotpoint}}
\put(546,297){\usebox{\plotpoint}}
\put(547,298){\usebox{\plotpoint}}
\put(548,300){\usebox{\plotpoint}}
\put(549,301){\usebox{\plotpoint}}
\put(550,302){\usebox{\plotpoint}}
\put(551,303){\usebox{\plotpoint}}
\put(552,305){\usebox{\plotpoint}}
\put(553,306){\usebox{\plotpoint}}
\put(554,307){\usebox{\plotpoint}}
\put(555,308){\usebox{\plotpoint}}
\put(556,310){\usebox{\plotpoint}}
\put(557,311){\usebox{\plotpoint}}
\put(558,312){\usebox{\plotpoint}}
\put(559,313){\usebox{\plotpoint}}
\put(560,314){\usebox{\plotpoint}}
\put(561,315){\usebox{\plotpoint}}
\put(562,316){\usebox{\plotpoint}}
\put(563,317){\usebox{\plotpoint}}
\put(564,319){\usebox{\plotpoint}}
\put(565,320){\usebox{\plotpoint}}
\put(566,321){\usebox{\plotpoint}}
\put(567,322){\usebox{\plotpoint}}
\put(568,324){\usebox{\plotpoint}}
\put(569,325){\usebox{\plotpoint}}
\put(570,326){\usebox{\plotpoint}}
\put(571,327){\usebox{\plotpoint}}
\put(572,328){\usebox{\plotpoint}}
\put(573,329){\usebox{\plotpoint}}
\put(574,330){\usebox{\plotpoint}}
\put(575,331){\usebox{\plotpoint}}
\put(576,332){\usebox{\plotpoint}}
\put(577,333){\usebox{\plotpoint}}
\put(578,334){\usebox{\plotpoint}}
\put(579,335){\usebox{\plotpoint}}
\put(580,336){\usebox{\plotpoint}}
\put(581,337){\usebox{\plotpoint}}
\put(582,338){\usebox{\plotpoint}}
\put(583,339){\usebox{\plotpoint}}
\put(584,340){\usebox{\plotpoint}}
\put(585,341){\usebox{\plotpoint}}
\put(586,342){\usebox{\plotpoint}}
\put(587,343){\usebox{\plotpoint}}
\put(589,344){\usebox{\plotpoint}}
\put(590,345){\usebox{\plotpoint}}
\put(591,346){\usebox{\plotpoint}}
\put(592,347){\usebox{\plotpoint}}
\put(593,348){\usebox{\plotpoint}}
\put(594,349){\usebox{\plotpoint}}
\put(595,350){\usebox{\plotpoint}}
\put(597,351){\usebox{\plotpoint}}
\put(598,352){\usebox{\plotpoint}}
\put(599,353){\usebox{\plotpoint}}
\put(601,354){\usebox{\plotpoint}}
\put(602,355){\usebox{\plotpoint}}
\put(603,356){\usebox{\plotpoint}}
\put(605,357){\usebox{\plotpoint}}
\put(606,358){\usebox{\plotpoint}}
\put(607,359){\usebox{\plotpoint}}
\put(610,360){\usebox{\plotpoint}}
\put(612,361){\usebox{\plotpoint}}
\put(614,362){\usebox{\plotpoint}}
\put(616,363){\usebox{\plotpoint}}
\put(617,364){\usebox{\plotpoint}}
\put(619,365){\usebox{\plotpoint}}
\put(621,366){\usebox{\plotpoint}}
\put(623,367){\usebox{\plotpoint}}
\put(625,368){\usebox{\plotpoint}}
\put(627,369){\usebox{\plotpoint}}
\put(631,370){\usebox{\plotpoint}}
\put(635,371){\usebox{\plotpoint}}
\put(639,372){\usebox{\plotpoint}}
\put(643,373){\usebox{\plotpoint}}
\put(647,374){\rule[-0.500pt]{1.927pt}{1.000pt}}
\sbox{\plotpoint}{\rule[-0.350pt]{0.700pt}{0.700pt}}%
\put(461,161){\rule{.1pt}{.1pt}}
\put(465,168){\rule{.1pt}{.1pt}}
\put(469,175){\rule{.1pt}{.1pt}}
\put(473,182){\rule{.1pt}{.1pt}}
\put(477,189){\rule{.1pt}{.1pt}}
\put(481,196){\rule{.1pt}{.1pt}}
\put(485,203){\rule{.1pt}{.1pt}}
\put(489,209){\rule{.1pt}{.1pt}}
\put(493,216){\rule{.1pt}{.1pt}}
\put(497,223){\rule{.1pt}{.1pt}}
\put(501,230){\rule{.1pt}{.1pt}}
\put(505,237){\rule{.1pt}{.1pt}}
\put(509,244){\rule{.1pt}{.1pt}}
\put(513,251){\rule{.1pt}{.1pt}}
\put(517,257){\rule{.1pt}{.1pt}}
\put(521,264){\rule{.1pt}{.1pt}}
\put(525,271){\rule{.1pt}{.1pt}}
\put(529,278){\rule{.1pt}{.1pt}}
\put(533,285){\rule{.1pt}{.1pt}}
\put(537,292){\rule{.1pt}{.1pt}}
\put(540,299){\rule{.1pt}{.1pt}}
\put(544,305){\rule{.1pt}{.1pt}}
\put(548,312){\rule{.1pt}{.1pt}}
\put(552,319){\rule{.1pt}{.1pt}}
\put(556,326){\rule{.1pt}{.1pt}}
\put(560,333){\rule{.1pt}{.1pt}}
\put(564,340){\rule{.1pt}{.1pt}}
\put(568,347){\rule{.1pt}{.1pt}}
\put(572,353){\rule{.1pt}{.1pt}}
\put(576,360){\rule{.1pt}{.1pt}}
\put(580,367){\rule{.1pt}{.1pt}}
\put(584,374){\rule{.1pt}{.1pt}}
\put(588,381){\rule{.1pt}{.1pt}}
\put(592,388){\rule{.1pt}{.1pt}}
\put(596,395){\rule{.1pt}{.1pt}}
\put(600,401){\rule{.1pt}{.1pt}}
\put(604,408){\rule{.1pt}{.1pt}}
\put(608,415){\rule{.1pt}{.1pt}}
\put(612,422){\rule{.1pt}{.1pt}}
\put(616,429){\rule{.1pt}{.1pt}}
\put(619,436){\rule{.1pt}{.1pt}}
\put(623,443){\rule{.1pt}{.1pt}}
\put(627,449){\rule{.1pt}{.1pt}}
\put(631,456){\rule{.1pt}{.1pt}}
\put(635,463){\rule{.1pt}{.1pt}}
\put(639,470){\rule{.1pt}{.1pt}}
\put(643,477){\rule{.1pt}{.1pt}}
\put(647,484){\rule{.1pt}{.1pt}}
\put(651,491){\rule{.1pt}{.1pt}}
\put(655,497){\rule{.1pt}{.1pt}}
\end{picture}
\hspace{-2.2cm} 
\setlength{\unitlength}{0.240900pt}
\ifx\plotpoint\undefined\newsavebox{\plotpoint}\fi
\sbox{\plotpoint}{\rule[-0.175pt]{0.350pt}{0.350pt}}%
\begin{picture}(719,660)(0,0)
\tenrm
\sbox{\plotpoint}{\rule[-0.175pt]{0.350pt}{0.350pt}}%
\put(264,158){\rule[-0.175pt]{94.192pt}{0.350pt}}
\put(655,158){\rule[-0.175pt]{0.350pt}{93.710pt}}
\put(264,547){\rule[-0.175pt]{94.192pt}{0.350pt}}
\put(199,352){\makebox(0,0)[l]{\shortstack{\Large $\Delta$}}}
\put(437,113){\makebox(0,0){\LARGE $t$}}
\put(557,498){\makebox(0,0)[l]{\Large (c)}}
\put(508,126){\makebox(0,0)[l]{{\large $t_c$} ($\sim B$)}}
\put(215,482){\makebox(0,0)[l]{$B$}}
\put(264,135){\makebox(0,0)[l]{\small 0}}
\put(231,168){\makebox(0,0)[l]{\small 0}}
\put(264,158){\rule[-0.175pt]{0.350pt}{93.710pt}}
\put(590,145){\line(0,1){26}}
\put(251,482){\line(1,0){26}}
\sbox{\plotpoint}{\rule[-0.500pt]{1.000pt}{1.000pt}}%
\put(521,234){\usebox{\plotpoint}}
\put(521,230){\usebox{\plotpoint}}
\put(522,228){\usebox{\plotpoint}}
\put(523,226){\usebox{\plotpoint}}
\put(524,224){\usebox{\plotpoint}}
\put(525,222){\usebox{\plotpoint}}
\put(526,219){\usebox{\plotpoint}}
\put(527,217){\usebox{\plotpoint}}
\put(528,215){\usebox{\plotpoint}}
\put(529,213){\usebox{\plotpoint}}
\put(530,212){\usebox{\plotpoint}}
\put(531,208){\usebox{\plotpoint}}
\put(532,206){\usebox{\plotpoint}}
\put(533,205){\usebox{\plotpoint}}
\put(534,203){\usebox{\plotpoint}}
\put(535,201){\usebox{\plotpoint}}
\put(536,198){\usebox{\plotpoint}}
\put(537,196){\usebox{\plotpoint}}
\put(538,195){\usebox{\plotpoint}}
\put(539,192){\usebox{\plotpoint}}
\put(540,190){\usebox{\plotpoint}}
\put(541,189){\usebox{\plotpoint}}
\put(542,189){\usebox{\plotpoint}}
\put(543,188){\usebox{\plotpoint}}
\put(544,184){\usebox{\plotpoint}}
\put(545,182){\usebox{\plotpoint}}
\put(546,181){\usebox{\plotpoint}}
\put(547,181){\usebox{\plotpoint}}
\put(548,180){\usebox{\plotpoint}}
\put(549,177){\usebox{\plotpoint}}
\put(550,177){\usebox{\plotpoint}}
\put(551,176){\usebox{\plotpoint}}
\put(552,173){\usebox{\plotpoint}}
\put(553,172){\usebox{\plotpoint}}
\put(554,172){\usebox{\plotpoint}}
\put(555,171){\usebox{\plotpoint}}
\put(556,170){\usebox{\plotpoint}}
\put(557,169){\usebox{\plotpoint}}
\put(558,168){\usebox{\plotpoint}}
\put(559,167){\usebox{\plotpoint}}
\put(560,166){\usebox{\plotpoint}}
\put(561,165){\usebox{\plotpoint}}
\put(562,164){\usebox{\plotpoint}}
\put(564,163){\usebox{\plotpoint}}
\put(565,162){\usebox{\plotpoint}}
\put(567,161){\usebox{\plotpoint}}
\put(568,160){\usebox{\plotpoint}}
\put(572,159){\usebox{\plotpoint}}
\put(575,158){\rule[-0.500pt]{3.132pt}{1.000pt}}
\put(264,482){\usebox{\plotpoint}}
\put(264,482){\usebox{\plotpoint}}
\put(267,481){\usebox{\plotpoint}}
\put(268,480){\usebox{\plotpoint}}
\put(269,479){\usebox{\plotpoint}}
\put(270,478){\usebox{\plotpoint}}
\put(272,477){\usebox{\plotpoint}}
\put(275,476){\usebox{\plotpoint}}
\put(276,475){\usebox{\plotpoint}}
\put(278,474){\usebox{\plotpoint}}
\put(279,473){\usebox{\plotpoint}}
\put(280,472){\usebox{\plotpoint}}
\put(281,471){\usebox{\plotpoint}}
\put(283,470){\usebox{\plotpoint}}
\put(286,469){\usebox{\plotpoint}}
\put(287,468){\usebox{\plotpoint}}
\put(289,467){\usebox{\plotpoint}}
\put(290,466){\usebox{\plotpoint}}
\put(291,465){\usebox{\plotpoint}}
\put(294,464){\usebox{\plotpoint}}
\put(295,463){\usebox{\plotpoint}}
\put(297,462){\usebox{\plotpoint}}
\put(298,461){\usebox{\plotpoint}}
\put(300,460){\usebox{\plotpoint}}
\put(301,459){\usebox{\plotpoint}}
\put(302,458){\usebox{\plotpoint}}
\put(305,457){\usebox{\plotpoint}}
\put(306,456){\usebox{\plotpoint}}
\put(308,455){\usebox{\plotpoint}}
\put(309,454){\usebox{\plotpoint}}
\put(311,453){\usebox{\plotpoint}}
\put(312,452){\usebox{\plotpoint}}
\put(313,451){\usebox{\plotpoint}}
\put(314,450){\usebox{\plotpoint}}
\put(316,449){\usebox{\plotpoint}}
\put(319,448){\usebox{\plotpoint}}
\put(320,447){\usebox{\plotpoint}}
\put(322,446){\usebox{\plotpoint}}
\put(323,445){\usebox{\plotpoint}}
\put(324,444){\usebox{\plotpoint}}
\put(325,443){\usebox{\plotpoint}}
\put(327,442){\usebox{\plotpoint}}
\put(328,441){\usebox{\plotpoint}}
\put(264,482){\rule{.1pt}{.1pt}}
\put(267,481){\rule{.1pt}{.1pt}}
\put(269,479){\rule{.1pt}{.1pt}}
\put(272,477){\rule{.1pt}{.1pt}}
\put(275,476){\rule{.1pt}{.1pt}}
\put(278,474){\rule{.1pt}{.1pt}}
\put(280,472){\rule{.1pt}{.1pt}}
\put(283,470){\rule{.1pt}{.1pt}}
\put(286,469){\rule{.1pt}{.1pt}}
\put(289,467){\rule{.1pt}{.1pt}}
\put(291,465){\rule{.1pt}{.1pt}}
\put(294,464){\rule{.1pt}{.1pt}}
\put(297,462){\rule{.1pt}{.1pt}}
\put(300,460){\rule{.1pt}{.1pt}}
\put(302,458){\rule{.1pt}{.1pt}}
\put(305,457){\rule{.1pt}{.1pt}}
\put(308,455){\rule{.1pt}{.1pt}}
\put(311,453){\rule{.1pt}{.1pt}}
\put(313,451){\rule{.1pt}{.1pt}}
\put(316,449){\rule{.1pt}{.1pt}}
\put(319,448){\rule{.1pt}{.1pt}}
\put(322,446){\rule{.1pt}{.1pt}}
\put(324,444){\rule{.1pt}{.1pt}}
\put(327,442){\rule{.1pt}{.1pt}}
\put(330,440){\rule{.1pt}{.1pt}}
\put(333,438){\rule{.1pt}{.1pt}}
\put(335,436){\rule{.1pt}{.1pt}}
\put(338,434){\rule{.1pt}{.1pt}}
\put(341,433){\rule{.1pt}{.1pt}}
\put(344,431){\rule{.1pt}{.1pt}}
\put(346,429){\rule{.1pt}{.1pt}}
\put(349,427){\rule{.1pt}{.1pt}}
\put(352,425){\rule{.1pt}{.1pt}}
\put(355,423){\rule{.1pt}{.1pt}}
\put(357,421){\rule{.1pt}{.1pt}}
\put(360,419){\rule{.1pt}{.1pt}}
\put(363,417){\rule{.1pt}{.1pt}}
\put(365,415){\rule{.1pt}{.1pt}}
\put(368,412){\rule{.1pt}{.1pt}}
\put(371,410){\rule{.1pt}{.1pt}}
\put(374,408){\rule{.1pt}{.1pt}}
\put(376,406){\rule{.1pt}{.1pt}}
\put(379,404){\rule{.1pt}{.1pt}}
\put(382,402){\rule{.1pt}{.1pt}}
\put(385,400){\rule{.1pt}{.1pt}}
\put(387,397){\rule{.1pt}{.1pt}}
\put(390,395){\rule{.1pt}{.1pt}}
\put(393,393){\rule{.1pt}{.1pt}}
\put(396,391){\rule{.1pt}{.1pt}}
\put(398,388){\rule{.1pt}{.1pt}}
\put(401,386){\rule{.1pt}{.1pt}}
\put(404,384){\rule{.1pt}{.1pt}}
\put(407,381){\rule{.1pt}{.1pt}}
\put(409,379){\rule{.1pt}{.1pt}}
\put(412,377){\rule{.1pt}{.1pt}}
\put(415,374){\rule{.1pt}{.1pt}}
\put(418,372){\rule{.1pt}{.1pt}}
\put(420,369){\rule{.1pt}{.1pt}}
\put(423,367){\rule{.1pt}{.1pt}}
\put(426,364){\rule{.1pt}{.1pt}}
\put(429,361){\rule{.1pt}{.1pt}}
\put(431,359){\rule{.1pt}{.1pt}}
\put(434,356){\rule{.1pt}{.1pt}}
\put(437,353){\rule{.1pt}{.1pt}}
\put(440,351){\rule{.1pt}{.1pt}}
\put(442,348){\rule{.1pt}{.1pt}}
\put(445,345){\rule{.1pt}{.1pt}}
\put(448,342){\rule{.1pt}{.1pt}}
\put(451,339){\rule{.1pt}{.1pt}}
\put(453,336){\rule{.1pt}{.1pt}}
\put(456,333){\rule{.1pt}{.1pt}}
\put(459,330){\rule{.1pt}{.1pt}}
\put(461,327){\rule{.1pt}{.1pt}}
\put(464,324){\rule{.1pt}{.1pt}}
\put(467,321){\rule{.1pt}{.1pt}}
\put(470,318){\rule{.1pt}{.1pt}}
\put(472,314){\rule{.1pt}{.1pt}}
\put(475,311){\rule{.1pt}{.1pt}}
\put(478,307){\rule{.1pt}{.1pt}}
\put(481,304){\rule{.1pt}{.1pt}}
\put(483,300){\rule{.1pt}{.1pt}}
\put(486,296){\rule{.1pt}{.1pt}}
\put(489,292){\rule{.1pt}{.1pt}}
\put(492,288){\rule{.1pt}{.1pt}}
\put(494,284){\rule{.1pt}{.1pt}}
\put(497,280){\rule{.1pt}{.1pt}}
\put(500,275){\rule{.1pt}{.1pt}}
\put(503,271){\rule{.1pt}{.1pt}}
\put(505,266){\rule{.1pt}{.1pt}}
\put(508,261){\rule{.1pt}{.1pt}}
\put(511,256){\rule{.1pt}{.1pt}}
\put(514,250){\rule{.1pt}{.1pt}}
\put(516,244){\rule{.1pt}{.1pt}}
\put(519,238){\rule{.1pt}{.1pt}}
\put(521,234){\rule{.1pt}{.1pt}}
\put(522,230){\rule{.1pt}{.1pt}}
\put(524,226){\rule{.1pt}{.1pt}}
\put(526,222){\rule{.1pt}{.1pt}}
\put(527,219){\rule{.1pt}{.1pt}}
\put(529,215){\rule{.1pt}{.1pt}}
\put(531,212){\rule{.1pt}{.1pt}}
\put(532,208){\rule{.1pt}{.1pt}}
\put(534,205){\rule{.1pt}{.1pt}}
\put(536,201){\rule{.1pt}{.1pt}}
\put(537,198){\rule{.1pt}{.1pt}}
\put(539,195){\rule{.1pt}{.1pt}}
\put(540,192){\rule{.1pt}{.1pt}}
\put(542,189){\rule{.1pt}{.1pt}}
\put(544,187){\rule{.1pt}{.1pt}}
\put(545,184){\rule{.1pt}{.1pt}}
\put(547,181){\rule{.1pt}{.1pt}}
\put(549,179){\rule{.1pt}{.1pt}}
\put(550,177){\rule{.1pt}{.1pt}}
\put(552,175){\rule{.1pt}{.1pt}}
\put(554,172){\rule{.1pt}{.1pt}}
\put(555,171){\rule{.1pt}{.1pt}}
\put(557,169){\rule{.1pt}{.1pt}}
\put(559,167){\rule{.1pt}{.1pt}}
\put(560,166){\rule{.1pt}{.1pt}}
\put(562,164){\rule{.1pt}{.1pt}}
\put(564,163){\rule{.1pt}{.1pt}}
\put(565,162){\rule{.1pt}{.1pt}}
\put(567,161){\rule{.1pt}{.1pt}}
\put(568,160){\rule{.1pt}{.1pt}}
\put(570,160){\rule{.1pt}{.1pt}}
\put(572,159){\rule{.1pt}{.1pt}}
\put(573,159){\rule{.1pt}{.1pt}}
\put(575,158){\rule{.1pt}{.1pt}}
\put(577,158){\rule{.1pt}{.1pt}}
\put(578,158){\rule{.1pt}{.1pt}}
\put(580,158){\rule{.1pt}{.1pt}}
\put(582,158){\rule{.1pt}{.1pt}}
\put(583,158){\rule{.1pt}{.1pt}}
\put(585,158){\rule{.1pt}{.1pt}}
\put(587,158){\rule{.1pt}{.1pt}}
\put(588,158){\rule{.1pt}{.1pt}}
\end{picture}
}
\vspace{-1.6cm}
\end{center}
\caption{Dispersion relation for the elementary excitations for the 1D
Hamiltonian, (a) for $t\! \ll\! B$, and (b) for $t\! \gg\! B$. (c) shows the
dependence of the gap $\Delta$ on $t$. $W$, $a$, and $v_s$ stand for
band width, lattice parameter, and sound velocity, respectively.}
\label{fig:gap}
\vspace{-6pt}
\end{figure}
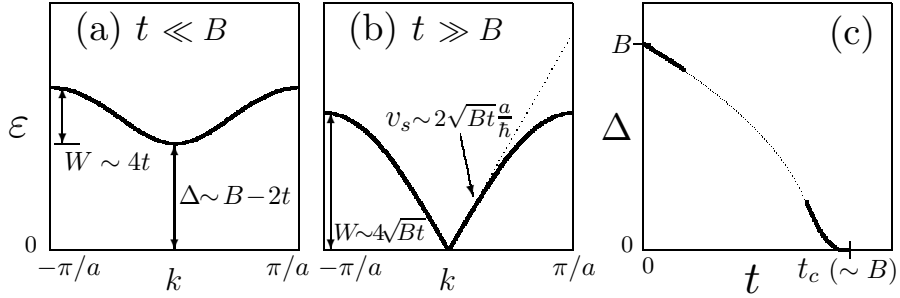

\section{Phase diagram}

\vspace{-2pt}

For $C\!=\! 0$ we have a one parameter ($\tilde {t}\! \equiv\! t/B$) problem.
The zero-temperature $d$-dimensional quantum Hamiltonian lies in the
universality class of the classical $(d+1)$-dimensional $XY$
model.\cite{jose} This model exhibits a phase transition 
at some critical $\tilde {t}_c$ of the order of 1. It is an 
order-disorder transition for $d\! >1\! $, implying a symmetry breaking in 
$\phi$ for the ordered phase ($\tilde {t}>\tilde {t}_c$), with 
its associated long-range order and gapless Goldstone excitations. 
For $d\! =\! 1$ the transition is of the Kosterlitz-Thouless kind. The 
$\tilde {t}>\tilde {t}_c$ ``quasi-ordered" phase displays
algebraic order ($\langle e^{i(\phi_i-\phi_j)}\rangle \sim 1/|i-j|^\eta$,
with $\eta \leq 1/4$). The disordered phase shows an exponential decay
of $\langle e^{i(\phi_i-\phi_j)}\rangle$, with gapped excitations.
For $d\! =\! 1$, near the transition, the gap closes like 
$\Delta\!\sim\! 1/\xi \!\sim\! e^{-K/\sqrt{\tilde t_c - \tilde t}}$, $\xi$
being the correlation length [Fig.2(c)].

The limits help to understand the nature of the phases.  For $\tilde 
t \! \ll \! 1$, the basis of independent-rotor eigenstates provides a 
natural language. The interaction term acts like a raising-lowering
operator on the rotor quantum numbers $m_i$. The original ${\cal H}$
can be mapped\cite{we} into a Hamiltonian of bosonic particles and 
antiparticles 
($m$ positive or negative) with a chemical potential $\mu\! =\! B$,
a Hubbard-like repulsion $U\! =\! B$, and a hopping energy $t$.
Using a Bogoliubov transformation, the $\tilde t \! \ll \! 1$ limit can be 
solved yielding an independent-rotor-like ground state with quantum 
fluctuations of particle-antiparticle pairs, and rotational excitations 
(rotatons, dressed rotor excitations), with $\varepsilon_k \! = \! B 
\sqrt{1\! - \! 4\tilde t\cos ka}$
[Fig.2(a)]. For $\tilde t \! \gg \! 1$, a harmonic approximation of the 
interaction term yields gapless spin-wave-like excitations [Fig.2(b)]. 

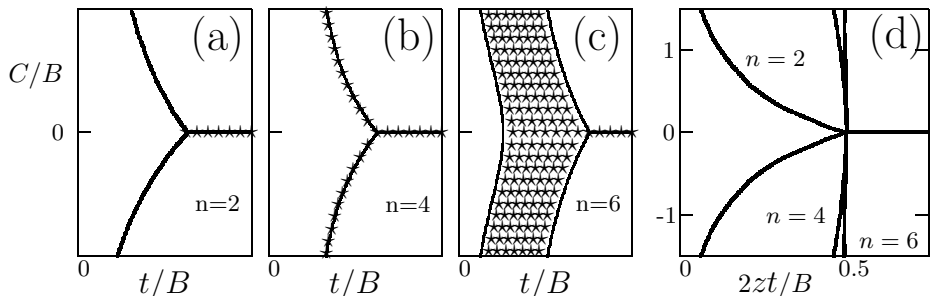
\begin{figure}
\begin{center}
\mbox{ \hspace {-1.7cm} 
\setlength{\unitlength}{0.240900pt}
\ifx\plotpoint\undefined\newsavebox{\plotpoint}\fi
\sbox{\plotpoint}{\rule[-0.175pt]{0.350pt}{0.350pt}}%
\begin{picture}(600,660)(0,0)
\tenrm
\sbox{\plotpoint}{\rule[-0.175pt]{0.350pt}{0.350pt}}%
\put(264,353){\rule[-0.175pt]{4.818pt}{0.350pt}}
\put(242,353){\makebox(0,0)[r]{\small $0$}}
\put(516,353){\rule[-0.175pt]{4.818pt}{0.350pt}}
\put(264,158){\rule[-0.175pt]{65.525pt}{0.350pt}}
\put(536,158){\rule[-0.175pt]{0.350pt}{93.710pt}}
\put(264,547){\rule[-0.175pt]{65.525pt}{0.350pt}}
\put(155,442){\makebox(0,0)[l]{\shortstack{$C/B$}}}
\put(400,113){\makebox(0,0){\large $t/B$}}
\put(264,139){\makebox(0,0)[l]{\small 0}}
\put(448,241){\makebox(0,0)[l]{\small n=2}}
\put(441,501){\makebox(0,0)[l]{\LARGE (a)}}
\put(264,158){\rule[-0.175pt]{0.350pt}{93.710pt}}
\sbox{\plotpoint}{\rule[-0.500pt]{1.000pt}{1.000pt}}%
\put(324,158){\rule[-0.500pt]{1.000pt}{1.124pt}}
\put(325,162){\usebox{\plotpoint}}
\put(326,165){\usebox{\plotpoint}}
\put(327,168){\usebox{\plotpoint}}
\put(328,170){\usebox{\plotpoint}}
\put(329,173){\usebox{\plotpoint}}
\put(330,176){\usebox{\plotpoint}}
\put(331,179){\usebox{\plotpoint}}
\put(332,182){\usebox{\plotpoint}}
\put(333,185){\usebox{\plotpoint}}
\put(334,187){\usebox{\plotpoint}}
\put(335,190){\usebox{\plotpoint}}
\put(336,193){\usebox{\plotpoint}}
\put(337,195){\usebox{\plotpoint}}
\put(338,198){\usebox{\plotpoint}}
\put(339,201){\usebox{\plotpoint}}
\put(340,203){\usebox{\plotpoint}}
\put(341,206){\usebox{\plotpoint}}
\put(342,209){\usebox{\plotpoint}}
\put(343,211){\usebox{\plotpoint}}
\put(344,213){\usebox{\plotpoint}}
\put(345,215){\usebox{\plotpoint}}
\put(346,217){\usebox{\plotpoint}}
\put(347,219){\usebox{\plotpoint}}
\put(348,221){\usebox{\plotpoint}}
\put(349,223){\usebox{\plotpoint}}
\put(350,226){\usebox{\plotpoint}}
\put(351,228){\usebox{\plotpoint}}
\put(352,230){\usebox{\plotpoint}}
\put(353,232){\usebox{\plotpoint}}
\put(354,234){\usebox{\plotpoint}}
\put(355,236){\usebox{\plotpoint}}
\put(356,238){\usebox{\plotpoint}}
\put(357,240){\usebox{\plotpoint}}
\put(358,242){\usebox{\plotpoint}}
\put(359,244){\usebox{\plotpoint}}
\put(360,246){\usebox{\plotpoint}}
\put(361,248){\usebox{\plotpoint}}
\put(362,250){\usebox{\plotpoint}}
\put(363,252){\usebox{\plotpoint}}
\put(364,254){\usebox{\plotpoint}}
\put(365,256){\usebox{\plotpoint}}
\put(366,258){\usebox{\plotpoint}}
\put(367,259){\usebox{\plotpoint}}
\put(368,261){\usebox{\plotpoint}}
\put(369,262){\usebox{\plotpoint}}
\put(370,264){\usebox{\plotpoint}}
\put(371,266){\usebox{\plotpoint}}
\put(372,267){\usebox{\plotpoint}}
\put(373,269){\usebox{\plotpoint}}
\put(374,271){\usebox{\plotpoint}}
\put(375,273){\usebox{\plotpoint}}
\put(376,274){\usebox{\plotpoint}}
\put(377,276){\usebox{\plotpoint}}
\put(378,278){\usebox{\plotpoint}}
\put(379,279){\usebox{\plotpoint}}
\put(380,281){\usebox{\plotpoint}}
\put(381,283){\usebox{\plotpoint}}
\put(382,284){\usebox{\plotpoint}}
\put(383,286){\usebox{\plotpoint}}
\put(384,287){\usebox{\plotpoint}}
\put(385,289){\usebox{\plotpoint}}
\put(386,291){\usebox{\plotpoint}}
\put(387,292){\usebox{\plotpoint}}
\put(388,293){\usebox{\plotpoint}}
\put(389,295){\usebox{\plotpoint}}
\put(390,296){\usebox{\plotpoint}}
\put(391,297){\usebox{\plotpoint}}
\put(392,299){\usebox{\plotpoint}}
\put(393,300){\usebox{\plotpoint}}
\put(394,302){\usebox{\plotpoint}}
\put(395,303){\usebox{\plotpoint}}
\put(396,305){\usebox{\plotpoint}}
\put(397,307){\usebox{\plotpoint}}
\put(398,308){\usebox{\plotpoint}}
\put(399,310){\usebox{\plotpoint}}
\put(400,311){\usebox{\plotpoint}}
\put(401,313){\usebox{\plotpoint}}
\put(402,314){\usebox{\plotpoint}}
\put(403,316){\usebox{\plotpoint}}
\put(404,317){\usebox{\plotpoint}}
\put(405,318){\usebox{\plotpoint}}
\put(406,319){\usebox{\plotpoint}}
\put(407,320){\usebox{\plotpoint}}
\put(408,321){\usebox{\plotpoint}}
\put(409,322){\usebox{\plotpoint}}
\put(410,323){\usebox{\plotpoint}}
\put(411,325){\usebox{\plotpoint}}
\put(412,326){\usebox{\plotpoint}}
\put(413,328){\usebox{\plotpoint}}
\put(414,329){\usebox{\plotpoint}}
\put(415,330){\usebox{\plotpoint}}
\put(416,332){\usebox{\plotpoint}}
\put(417,333){\usebox{\plotpoint}}
\put(418,334){\usebox{\plotpoint}}
\put(419,335){\usebox{\plotpoint}}
\put(420,336){\usebox{\plotpoint}}
\put(421,337){\usebox{\plotpoint}}
\put(422,338){\usebox{\plotpoint}}
\put(423,339){\usebox{\plotpoint}}
\put(424,341){\usebox{\plotpoint}}
\put(425,342){\usebox{\plotpoint}}
\put(426,343){\usebox{\plotpoint}}
\put(427,344){\usebox{\plotpoint}}
\put(428,345){\usebox{\plotpoint}}
\put(429,346){\usebox{\plotpoint}}
\put(430,347){\usebox{\plotpoint}}
\put(431,349){\usebox{\plotpoint}}
\put(432,350){\usebox{\plotpoint}}
\put(433,351){\usebox{\plotpoint}}
\put(434,353){\rule[-0.500pt]{24.572pt}{1.000pt}}
\put(434,353){\rule[-0.500pt]{24.572pt}{1.000pt}}
\put(434,353){\usebox{\plotpoint}}
\put(433,354){\usebox{\plotpoint}}
\put(432,355){\usebox{\plotpoint}}
\put(431,357){\usebox{\plotpoint}}
\put(430,358){\usebox{\plotpoint}}
\put(429,359){\usebox{\plotpoint}}
\put(428,361){\usebox{\plotpoint}}
\put(427,362){\usebox{\plotpoint}}
\put(426,363){\usebox{\plotpoint}}
\put(425,365){\usebox{\plotpoint}}
\put(424,366){\usebox{\plotpoint}}
\put(423,368){\usebox{\plotpoint}}
\put(422,369){\usebox{\plotpoint}}
\put(421,371){\usebox{\plotpoint}}
\put(420,373){\usebox{\plotpoint}}
\put(419,374){\usebox{\plotpoint}}
\put(418,375){\usebox{\plotpoint}}
\put(417,377){\usebox{\plotpoint}}
\put(416,378){\usebox{\plotpoint}}
\put(415,379){\usebox{\plotpoint}}
\put(414,381){\usebox{\plotpoint}}
\put(413,382){\usebox{\plotpoint}}
\put(412,384){\usebox{\plotpoint}}
\put(411,385){\usebox{\plotpoint}}
\put(410,387){\usebox{\plotpoint}}
\put(409,389){\usebox{\plotpoint}}
\put(408,391){\usebox{\plotpoint}}
\put(407,393){\usebox{\plotpoint}}
\put(406,395){\usebox{\plotpoint}}
\put(405,398){\usebox{\plotpoint}}
\put(404,399){\usebox{\plotpoint}}
\put(403,401){\usebox{\plotpoint}}
\put(402,402){\usebox{\plotpoint}}
\put(401,404){\usebox{\plotpoint}}
\put(400,406){\usebox{\plotpoint}}
\put(399,408){\usebox{\plotpoint}}
\put(398,410){\usebox{\plotpoint}}
\put(397,412){\usebox{\plotpoint}}
\put(396,414){\usebox{\plotpoint}}
\put(395,415){\usebox{\plotpoint}}
\put(394,417){\usebox{\plotpoint}}
\put(393,418){\usebox{\plotpoint}}
\put(392,420){\usebox{\plotpoint}}
\put(391,422){\usebox{\plotpoint}}
\put(390,424){\usebox{\plotpoint}}
\put(389,426){\usebox{\plotpoint}}
\put(388,428){\usebox{\plotpoint}}
\put(387,430){\usebox{\plotpoint}}
\put(386,432){\usebox{\plotpoint}}
\put(385,434){\usebox{\plotpoint}}
\put(384,436){\usebox{\plotpoint}}
\put(383,439){\usebox{\plotpoint}}
\put(382,441){\usebox{\plotpoint}}
\put(381,444){\usebox{\plotpoint}}
\put(380,446){\usebox{\plotpoint}}
\put(379,449){\usebox{\plotpoint}}
\put(378,451){\usebox{\plotpoint}}
\put(377,453){\usebox{\plotpoint}}
\put(376,455){\usebox{\plotpoint}}
\put(375,457){\usebox{\plotpoint}}
\put(374,460){\usebox{\plotpoint}}
\put(373,462){\usebox{\plotpoint}}
\put(372,465){\usebox{\plotpoint}}
\put(371,468){\usebox{\plotpoint}}
\put(370,470){\usebox{\plotpoint}}
\put(369,473){\usebox{\plotpoint}}
\put(368,476){\usebox{\plotpoint}}
\put(367,478){\usebox{\plotpoint}}
\put(366,482){\usebox{\plotpoint}}
\put(365,485){\usebox{\plotpoint}}
\put(364,488){\usebox{\plotpoint}}
\put(363,490){\usebox{\plotpoint}}
\put(362,493){\usebox{\plotpoint}}
\put(361,495){\usebox{\plotpoint}}
\put(360,498){\usebox{\plotpoint}}
\put(359,501){\usebox{\plotpoint}}
\put(358,503){\usebox{\plotpoint}}
\put(357,506){\usebox{\plotpoint}}
\put(356,509){\usebox{\plotpoint}}
\put(355,511){\usebox{\plotpoint}}
\put(354,516){\usebox{\plotpoint}}
\put(353,520){\rule[-0.500pt]{1.000pt}{1.084pt}}
\put(352,524){\rule[-0.500pt]{1.000pt}{1.084pt}}
\put(351,529){\usebox{\plotpoint}}
\put(350,531){\usebox{\plotpoint}}
\put(349,534){\usebox{\plotpoint}}
\put(348,537){\usebox{\plotpoint}}
\put(347,541){\usebox{\plotpoint}}
\put(346,545){\usebox{\plotpoint}}
\sbox{\plotpoint}{\rule[-0.350pt]{0.700pt}{0.700pt}}%
\put(434,353){\makebox(0,0){$\star$}}
\put(451,353){\makebox(0,0){$\star$}}
\put(468,353){\makebox(0,0){$\star$}}
\put(485,353){\makebox(0,0){$\star$}}
\put(502,353){\makebox(0,0){$\star$}}
\put(519,353){\makebox(0,0){$\star$}}
\put(536,353){\makebox(0,0){$\star$}}
\end{picture}
\hspace{-2.9cm} 
\setlength{\unitlength}{0.240900pt}
\ifx\plotpoint\undefined\newsavebox{\plotpoint}\fi
\sbox{\plotpoint}{\rule[-0.175pt]{0.350pt}{0.350pt}}%
\begin{picture}(600,660)(0,0)
\tenrm
\sbox{\plotpoint}{\rule[-0.175pt]{0.350pt}{0.350pt}}%
\put(264,353){\rule[-0.175pt]{4.818pt}{0.350pt}}
\put(516,353){\rule[-0.175pt]{4.818pt}{0.350pt}}
\put(264,158){\rule[-0.175pt]{65.525pt}{0.350pt}}
\put(536,158){\rule[-0.175pt]{0.350pt}{93.710pt}}
\put(264,547){\rule[-0.175pt]{65.525pt}{0.350pt}}
\put(400,113){\makebox(0,0){\large $t/B$}}
\put(264,139){\makebox(0,0)[l]{\small 0}}
\put(448,240){\makebox(0,0)[l]{\small n=4}}
\put(441,500){\makebox(0,0)[l]{\LARGE (b)}}
\put(264,158){\rule[-0.175pt]{0.350pt}{93.710pt}}
\sbox{\plotpoint}{\rule[-0.350pt]{0.700pt}{0.700pt}}%
\put(354,158){\rule[-0.350pt]{0.700pt}{2.168pt}}
\put(355,167){\rule[-0.350pt]{0.700pt}{1.204pt}}
\put(356,172){\rule[-0.350pt]{0.700pt}{1.445pt}}
\put(357,178){\rule[-0.350pt]{0.700pt}{1.204pt}}
\put(358,183){\rule[-0.350pt]{0.700pt}{0.723pt}}
\put(359,186){\rule[-0.350pt]{0.700pt}{0.723pt}}
\put(360,189){\rule[-0.350pt]{0.700pt}{1.204pt}}
\put(361,194){\rule[-0.350pt]{0.700pt}{1.445pt}}
\put(362,200){\usebox{\plotpoint}}
\put(363,202){\usebox{\plotpoint}}
\put(364,205){\rule[-0.350pt]{0.700pt}{1.445pt}}
\put(365,211){\rule[-0.350pt]{0.700pt}{0.723pt}}
\put(366,214){\rule[-0.350pt]{0.700pt}{0.723pt}}
\put(367,217){\rule[-0.350pt]{0.700pt}{1.204pt}}
\put(368,222){\rule[-0.350pt]{0.700pt}{0.723pt}}
\put(369,225){\rule[-0.350pt]{0.700pt}{0.723pt}}
\put(370,228){\rule[-0.350pt]{0.700pt}{1.204pt}}
\put(371,233){\rule[-0.350pt]{0.700pt}{0.723pt}}
\put(372,236){\rule[-0.350pt]{0.700pt}{0.723pt}}
\put(373,239){\usebox{\plotpoint}}
\put(374,241){\usebox{\plotpoint}}
\put(375,244){\rule[-0.350pt]{0.700pt}{0.723pt}}
\put(376,247){\rule[-0.350pt]{0.700pt}{0.723pt}}
\put(377,250){\usebox{\plotpoint}}
\put(378,252){\usebox{\plotpoint}}
\put(379,255){\rule[-0.350pt]{0.700pt}{0.723pt}}
\put(380,258){\rule[-0.350pt]{0.700pt}{0.723pt}}
\put(381,261){\usebox{\plotpoint}}
\put(382,263){\usebox{\plotpoint}}
\put(383,266){\usebox{\plotpoint}}
\put(384,268){\usebox{\plotpoint}}
\put(385,270){\usebox{\plotpoint}}
\put(386,272){\rule[-0.350pt]{0.700pt}{0.723pt}}
\put(387,275){\rule[-0.350pt]{0.700pt}{0.723pt}}
\put(388,278){\usebox{\plotpoint}}
\put(389,279){\usebox{\plotpoint}}
\put(390,281){\usebox{\plotpoint}}
\put(391,282){\rule[-0.350pt]{0.700pt}{0.723pt}}
\put(392,286){\rule[-0.350pt]{0.700pt}{0.723pt}}
\put(393,289){\usebox{\plotpoint}}
\put(394,290){\usebox{\plotpoint}}
\put(395,292){\usebox{\plotpoint}}
\put(396,293){\usebox{\plotpoint}}
\put(397,296){\usebox{\plotpoint}}
\put(398,298){\usebox{\plotpoint}}
\put(399,300){\usebox{\plotpoint}}
\put(400,301){\usebox{\plotpoint}}
\put(401,303){\usebox{\plotpoint}}
\put(402,304){\usebox{\plotpoint}}
\put(403,307){\usebox{\plotpoint}}
\put(404,309){\usebox{\plotpoint}}
\put(405,311){\usebox{\plotpoint}}
\put(406,312){\usebox{\plotpoint}}
\put(407,313){\usebox{\plotpoint}}
\put(408,314){\usebox{\plotpoint}}
\put(409,316){\usebox{\plotpoint}}
\put(410,318){\usebox{\plotpoint}}
\put(411,320){\usebox{\plotpoint}}
\put(412,322){\usebox{\plotpoint}}
\put(413,323){\usebox{\plotpoint}}
\put(414,325){\usebox{\plotpoint}}
\put(415,326){\usebox{\plotpoint}}
\put(416,328){\usebox{\plotpoint}}
\put(417,329){\usebox{\plotpoint}}
\put(418,331){\usebox{\plotpoint}}
\put(419,332){\usebox{\plotpoint}}
\put(420,334){\usebox{\plotpoint}}
\put(421,336){\usebox{\plotpoint}}
\put(422,337){\usebox{\plotpoint}}
\put(423,339){\usebox{\plotpoint}}
\put(424,340){\usebox{\plotpoint}}
\put(425,341){\usebox{\plotpoint}}
\put(426,342){\usebox{\plotpoint}}
\put(427,344){\usebox{\plotpoint}}
\put(428,345){\usebox{\plotpoint}}
\put(429,346){\usebox{\plotpoint}}
\put(430,347){\usebox{\plotpoint}}
\put(431,348){\usebox{\plotpoint}}
\put(432,350){\usebox{\plotpoint}}
\put(433,351){\usebox{\plotpoint}}
\put(434,353){\rule[-0.350pt]{24.572pt}{0.700pt}}
\put(433,353){\rule[-0.350pt]{24.813pt}{0.700pt}}
\put(432,354){\usebox{\plotpoint}}
\put(432,355){\usebox{\plotpoint}}
\put(431,356){\usebox{\plotpoint}}
\put(430,357){\usebox{\plotpoint}}
\put(429,358){\usebox{\plotpoint}}
\put(428,359){\usebox{\plotpoint}}
\put(427,361){\usebox{\plotpoint}}
\put(426,362){\usebox{\plotpoint}}
\put(425,363){\usebox{\plotpoint}}
\put(424,364){\usebox{\plotpoint}}
\put(423,366){\usebox{\plotpoint}}
\put(422,367){\usebox{\plotpoint}}
\put(421,369){\usebox{\plotpoint}}
\put(420,370){\usebox{\plotpoint}}
\put(419,372){\usebox{\plotpoint}}
\put(418,373){\usebox{\plotpoint}}
\put(417,375){\usebox{\plotpoint}}
\put(416,376){\usebox{\plotpoint}}
\put(415,378){\usebox{\plotpoint}}
\put(414,380){\usebox{\plotpoint}}
\put(413,381){\usebox{\plotpoint}}
\put(412,383){\usebox{\plotpoint}}
\put(411,385){\usebox{\plotpoint}}
\put(410,387){\usebox{\plotpoint}}
\put(409,389){\usebox{\plotpoint}}
\put(408,390){\usebox{\plotpoint}}
\put(407,391){\usebox{\plotpoint}}
\put(406,392){\usebox{\plotpoint}}
\put(405,394){\usebox{\plotpoint}}
\put(404,396){\usebox{\plotpoint}}
\put(403,398){\usebox{\plotpoint}}
\put(402,400){\usebox{\plotpoint}}
\put(401,401){\usebox{\plotpoint}}
\put(400,403){\usebox{\plotpoint}}
\put(399,404){\usebox{\plotpoint}}
\put(398,407){\usebox{\plotpoint}}
\put(397,409){\usebox{\plotpoint}}
\put(396,411){\usebox{\plotpoint}}
\put(395,412){\usebox{\plotpoint}}
\put(394,414){\usebox{\plotpoint}}
\put(393,415){\rule[-0.350pt]{0.700pt}{0.723pt}}
\put(392,419){\rule[-0.350pt]{0.700pt}{0.723pt}}
\put(391,422){\usebox{\plotpoint}}
\put(390,423){\usebox{\plotpoint}}
\put(389,425){\usebox{\plotpoint}}
\put(388,426){\rule[-0.350pt]{0.700pt}{0.723pt}}
\put(387,430){\rule[-0.350pt]{0.700pt}{0.723pt}}
\put(386,433){\usebox{\plotpoint}}
\put(385,435){\usebox{\plotpoint}}
\put(384,437){\usebox{\plotpoint}}
\put(383,439){\usebox{\plotpoint}}
\put(382,441){\usebox{\plotpoint}}
\put(381,444){\rule[-0.350pt]{0.700pt}{0.723pt}}
\put(380,447){\rule[-0.350pt]{0.700pt}{0.723pt}}
\put(379,450){\usebox{\plotpoint}}
\put(378,452){\usebox{\plotpoint}}
\put(377,455){\rule[-0.350pt]{0.700pt}{0.723pt}}
\put(376,458){\rule[-0.350pt]{0.700pt}{0.723pt}}
\put(375,461){\usebox{\plotpoint}}
\put(374,463){\usebox{\plotpoint}}
\put(373,466){\rule[-0.350pt]{0.700pt}{0.723pt}}
\put(372,469){\rule[-0.350pt]{0.700pt}{0.723pt}}
\put(371,472){\rule[-0.350pt]{0.700pt}{1.204pt}}
\put(370,477){\rule[-0.350pt]{0.700pt}{0.723pt}}
\put(369,480){\rule[-0.350pt]{0.700pt}{0.723pt}}
\put(368,483){\rule[-0.350pt]{0.700pt}{1.204pt}}
\put(367,488){\rule[-0.350pt]{0.700pt}{0.723pt}}
\put(366,491){\rule[-0.350pt]{0.700pt}{0.723pt}}
\put(365,494){\rule[-0.350pt]{0.700pt}{1.445pt}}
\put(364,500){\usebox{\plotpoint}}
\put(363,502){\usebox{\plotpoint}}
\put(362,505){\rule[-0.350pt]{0.700pt}{1.445pt}}
\put(361,511){\rule[-0.350pt]{0.700pt}{1.204pt}}
\put(360,516){\rule[-0.350pt]{0.700pt}{0.723pt}}
\put(359,519){\rule[-0.350pt]{0.700pt}{0.723pt}}
\put(358,522){\rule[-0.350pt]{0.700pt}{1.204pt}}
\put(357,527){\rule[-0.350pt]{0.700pt}{1.445pt}}
\put(356,533){\rule[-0.350pt]{0.700pt}{1.204pt}}
\put(355,538){\rule[-0.350pt]{0.700pt}{1.445pt}}
\put(354,544){\rule[-0.350pt]{0.700pt}{0.723pt}}
\put(355,164){\makebox(0,0){$\star$}}
\put(358,183){\makebox(0,0){$\star$}}
\put(363,202){\makebox(0,0){$\star$}}
\put(368,221){\makebox(0,0){$\star$}}
\put(373,240){\makebox(0,0){$\star$}}
\put(380,258){\makebox(0,0){$\star$}}
\put(388,277){\makebox(0,0){$\star$}}
\put(397,296){\makebox(0,0){$\star$}}
\put(408,315){\makebox(0,0){$\star$}}
\put(420,334){\makebox(0,0){$\star$}}
\put(434,353){\makebox(0,0){$\star$}}
\put(451,353){\makebox(0,0){$\star$}}
\put(468,353){\makebox(0,0){$\star$}}
\put(485,353){\makebox(0,0){$\star$}}
\put(502,353){\makebox(0,0){$\star$}}
\put(519,353){\makebox(0,0){$\star$}}
\put(536,353){\makebox(0,0){$\star$}}
\put(420,371){\makebox(0,0){$\star$}}
\put(408,390){\makebox(0,0){$\star$}}
\put(397,409){\makebox(0,0){$\star$}}
\put(388,428){\makebox(0,0){$\star$}}
\put(380,447){\makebox(0,0){$\star$}}
\put(373,465){\makebox(0,0){$\star$}}
\put(368,484){\makebox(0,0){$\star$}}
\put(363,503){\makebox(0,0){$\star$}}
\put(358,522){\makebox(0,0){$\star$}}
\put(355,541){\makebox(0,0){$\star$}}
\end{picture}
\hspace{-2.9cm} 
\setlength{\unitlength}{0.240900pt}
\ifx\plotpoint\undefined\newsavebox{\plotpoint}\fi
\sbox{\plotpoint}{\rule[-0.175pt]{0.350pt}{0.350pt}}%
\begin{picture}(600,660)(0,0)
\tenrm
\sbox{\plotpoint}{\rule[-0.175pt]{0.350pt}{0.350pt}}%
\put(264,353){\rule[-0.175pt]{4.818pt}{0.350pt}}
\put(516,353){\rule[-0.175pt]{4.818pt}{0.350pt}}
\put(264,158){\rule[-0.175pt]{65.525pt}{0.350pt}}
\put(536,158){\rule[-0.175pt]{0.350pt}{93.710pt}}
\put(264,547){\rule[-0.175pt]{65.525pt}{0.350pt}}
\put(400,113){\makebox(0,0){\large $t/B$}}
\put(264,139){\makebox(0,0)[l]{\small 0}}
\put(448,241){\makebox(0,0)[l]{\small n=6}}
\put(441,501){\makebox(0,0)[l]{\LARGE (c)}}
\put(264,158){\rule[-0.175pt]{0.350pt}{93.710pt}}
\sbox{\plotpoint}{\rule[-0.350pt]{0.700pt}{0.700pt}}%
\put(296,158){\usebox{\plotpoint}}
\put(297,159){\rule[-0.350pt]{0.700pt}{1.445pt}}
\put(298,165){\rule[-0.350pt]{0.700pt}{1.445pt}}
\put(299,171){\rule[-0.350pt]{0.700pt}{0.964pt}}
\put(300,175){\rule[-0.350pt]{0.700pt}{1.686pt}}
\put(301,182){\rule[-0.350pt]{0.700pt}{0.964pt}}
\put(302,186){\rule[-0.350pt]{0.700pt}{1.445pt}}
\put(303,192){\rule[-0.350pt]{0.700pt}{0.964pt}}
\put(304,196){\rule[-0.350pt]{0.700pt}{0.964pt}}
\put(305,200){\rule[-0.350pt]{0.700pt}{1.445pt}}
\put(306,206){\rule[-0.350pt]{0.700pt}{0.964pt}}
\put(307,210){\rule[-0.350pt]{0.700pt}{0.964pt}}
\put(308,214){\rule[-0.350pt]{0.700pt}{0.964pt}}
\put(309,218){\rule[-0.350pt]{0.700pt}{0.964pt}}
\put(310,222){\rule[-0.350pt]{0.700pt}{1.204pt}}
\put(311,227){\rule[-0.350pt]{0.700pt}{1.445pt}}
\put(312,233){\rule[-0.350pt]{0.700pt}{0.964pt}}
\put(313,237){\rule[-0.350pt]{0.700pt}{0.964pt}}
\put(314,241){\rule[-0.350pt]{0.700pt}{0.964pt}}
\put(315,245){\rule[-0.350pt]{0.700pt}{0.964pt}}
\put(316,249){\rule[-0.350pt]{0.700pt}{0.964pt}}
\put(317,253){\rule[-0.350pt]{0.700pt}{0.964pt}}
\put(318,257){\rule[-0.350pt]{0.700pt}{0.964pt}}
\put(319,261){\rule[-0.350pt]{0.700pt}{0.964pt}}
\put(320,265){\rule[-0.350pt]{0.700pt}{1.204pt}}
\put(321,270){\rule[-0.350pt]{0.700pt}{0.964pt}}
\put(322,274){\rule[-0.350pt]{0.700pt}{1.445pt}}
\put(323,280){\rule[-0.350pt]{0.700pt}{0.964pt}}
\put(324,284){\rule[-0.350pt]{0.700pt}{0.964pt}}
\put(325,288){\rule[-0.350pt]{0.700pt}{1.445pt}}
\put(326,294){\rule[-0.350pt]{0.700pt}{0.964pt}}
\put(327,298){\rule[-0.350pt]{0.700pt}{1.445pt}}
\put(328,304){\rule[-0.350pt]{0.700pt}{1.686pt}}
\put(329,311){\rule[-0.350pt]{0.700pt}{1.445pt}}
\put(330,317){\rule[-0.350pt]{0.700pt}{1.927pt}}
\put(331,325){\rule[-0.350pt]{0.700pt}{2.891pt}}
\put(332,337){\rule[-0.350pt]{0.700pt}{7.950pt}}
\put(331,370){\rule[-0.350pt]{0.700pt}{2.891pt}}
\put(330,382){\rule[-0.350pt]{0.700pt}{1.927pt}}
\put(329,390){\rule[-0.350pt]{0.700pt}{1.686pt}}
\put(328,397){\rule[-0.350pt]{0.700pt}{1.445pt}}
\put(327,403){\rule[-0.350pt]{0.700pt}{1.445pt}}
\put(326,409){\rule[-0.350pt]{0.700pt}{0.964pt}}
\put(325,413){\rule[-0.350pt]{0.700pt}{1.445pt}}
\put(324,419){\rule[-0.350pt]{0.700pt}{0.964pt}}
\put(323,423){\rule[-0.350pt]{0.700pt}{0.964pt}}
\put(322,427){\rule[-0.350pt]{0.700pt}{1.445pt}}
\put(321,433){\rule[-0.350pt]{0.700pt}{0.964pt}}
\put(320,437){\rule[-0.350pt]{0.700pt}{1.204pt}}
\put(319,442){\rule[-0.350pt]{0.700pt}{0.964pt}}
\put(318,446){\rule[-0.350pt]{0.700pt}{0.964pt}}
\put(317,450){\rule[-0.350pt]{0.700pt}{0.964pt}}
\put(316,454){\rule[-0.350pt]{0.700pt}{0.964pt}}
\put(315,458){\rule[-0.350pt]{0.700pt}{0.964pt}}
\put(314,462){\rule[-0.350pt]{0.700pt}{0.964pt}}
\put(313,466){\rule[-0.350pt]{0.700pt}{0.964pt}}
\put(312,470){\rule[-0.350pt]{0.700pt}{0.964pt}}
\put(311,474){\rule[-0.350pt]{0.700pt}{1.445pt}}
\put(310,480){\rule[-0.350pt]{0.700pt}{1.204pt}}
\put(309,485){\rule[-0.350pt]{0.700pt}{0.964pt}}
\put(308,489){\rule[-0.350pt]{0.700pt}{0.964pt}}
\put(307,493){\rule[-0.350pt]{0.700pt}{0.964pt}}
\put(306,497){\rule[-0.350pt]{0.700pt}{0.964pt}}
\put(305,501){\rule[-0.350pt]{0.700pt}{1.445pt}}
\put(304,507){\rule[-0.350pt]{0.700pt}{0.964pt}}
\put(303,511){\rule[-0.350pt]{0.700pt}{0.964pt}}
\put(302,515){\rule[-0.350pt]{0.700pt}{1.445pt}}
\put(301,521){\rule[-0.350pt]{0.700pt}{1.204pt}}
\put(300,526){\rule[-0.350pt]{0.700pt}{1.445pt}}
\put(299,532){\rule[-0.350pt]{0.700pt}{0.964pt}}
\put(298,536){\rule[-0.350pt]{0.700pt}{1.445pt}}
\put(297,542){\rule[-0.350pt]{0.700pt}{0.964pt}}
\put(296,546){\usebox{\plotpoint}}
\put(402,158){\rule[-0.350pt]{0.700pt}{1.204pt}}
\put(403,163){\rule[-0.350pt]{0.700pt}{0.964pt}}
\put(404,167){\rule[-0.350pt]{0.700pt}{1.445pt}}
\put(405,173){\rule[-0.350pt]{0.700pt}{0.964pt}}
\put(406,177){\rule[-0.350pt]{0.700pt}{1.204pt}}
\put(407,182){\rule[-0.350pt]{0.700pt}{1.445pt}}
\put(408,188){\rule[-0.350pt]{0.700pt}{0.964pt}}
\put(409,192){\rule[-0.350pt]{0.700pt}{0.964pt}}
\put(410,196){\rule[-0.350pt]{0.700pt}{0.964pt}}
\put(411,200){\rule[-0.350pt]{0.700pt}{0.964pt}}
\put(412,204){\rule[-0.350pt]{0.700pt}{0.964pt}}
\put(413,208){\rule[-0.350pt]{0.700pt}{0.964pt}}
\put(414,212){\rule[-0.350pt]{0.700pt}{0.964pt}}
\put(415,216){\rule[-0.350pt]{0.700pt}{0.964pt}}
\put(416,220){\usebox{\plotpoint}}
\put(417,222){\rule[-0.350pt]{0.700pt}{1.204pt}}
\put(418,227){\rule[-0.350pt]{0.700pt}{0.964pt}}
\put(419,231){\rule[-0.350pt]{0.700pt}{0.964pt}}
\put(420,235){\usebox{\plotpoint}}
\put(421,237){\rule[-0.350pt]{0.700pt}{0.964pt}}
\put(422,241){\rule[-0.350pt]{0.700pt}{0.964pt}}
\put(423,245){\usebox{\plotpoint}}
\put(424,247){\rule[-0.350pt]{0.700pt}{0.964pt}}
\put(425,251){\usebox{\plotpoint}}
\put(426,253){\rule[-0.350pt]{0.700pt}{0.964pt}}
\put(427,257){\usebox{\plotpoint}}
\put(428,259){\rule[-0.350pt]{0.700pt}{0.964pt}}
\put(429,263){\usebox{\plotpoint}}
\put(430,265){\rule[-0.350pt]{0.700pt}{1.204pt}}
\put(431,270){\usebox{\plotpoint}}
\put(432,272){\usebox{\plotpoint}}
\put(433,274){\rule[-0.350pt]{0.700pt}{0.964pt}}
\put(434,278){\usebox{\plotpoint}}
\put(435,280){\usebox{\plotpoint}}
\put(436,282){\rule[-0.350pt]{0.700pt}{0.964pt}}
\put(437,286){\usebox{\plotpoint}}
\put(438,288){\usebox{\plotpoint}}
\put(439,290){\usebox{\plotpoint}}
\put(440,292){\rule[-0.350pt]{0.700pt}{0.964pt}}
\put(441,296){\usebox{\plotpoint}}
\put(442,298){\usebox{\plotpoint}}
\put(443,300){\usebox{\plotpoint}}
\put(444,302){\usebox{\plotpoint}}
\put(445,304){\usebox{\plotpoint}}
\put(446,306){\rule[-0.350pt]{0.700pt}{1.204pt}}
\put(447,311){\usebox{\plotpoint}}
\put(448,313){\usebox{\plotpoint}}
\put(449,315){\usebox{\plotpoint}}
\put(450,317){\usebox{\plotpoint}}
\put(451,319){\usebox{\plotpoint}}
\put(452,321){\usebox{\plotpoint}}
\put(453,323){\usebox{\plotpoint}}
\put(454,325){\usebox{\plotpoint}}
\put(455,327){\usebox{\plotpoint}}
\put(456,329){\usebox{\plotpoint}}
\put(457,331){\usebox{\plotpoint}}
\put(458,333){\usebox{\plotpoint}}
\put(459,335){\usebox{\plotpoint}}
\put(460,337){\usebox{\plotpoint}}
\put(461,339){\usebox{\plotpoint}}
\put(462,341){\usebox{\plotpoint}}
\put(463,343){\usebox{\plotpoint}}
\put(464,345){\usebox{\plotpoint}}
\put(465,347){\usebox{\plotpoint}}
\put(466,349){\usebox{\plotpoint}}
\put(467,351){\usebox{\plotpoint}}
\put(468,353){\rule[-0.350pt]{16.381pt}{0.700pt}}
\put(467,353){\rule[-0.350pt]{16.622pt}{0.700pt}}
\put(467,354){\usebox{\plotpoint}}
\put(466,356){\usebox{\plotpoint}}
\put(465,358){\usebox{\plotpoint}}
\put(464,360){\usebox{\plotpoint}}
\put(463,362){\usebox{\plotpoint}}
\put(462,364){\usebox{\plotpoint}}
\put(461,366){\usebox{\plotpoint}}
\put(460,368){\usebox{\plotpoint}}
\put(459,370){\usebox{\plotpoint}}
\put(458,372){\usebox{\plotpoint}}
\put(457,374){\usebox{\plotpoint}}
\put(456,376){\usebox{\plotpoint}}
\put(455,378){\usebox{\plotpoint}}
\put(454,380){\usebox{\plotpoint}}
\put(453,382){\usebox{\plotpoint}}
\put(452,384){\usebox{\plotpoint}}
\put(451,386){\usebox{\plotpoint}}
\put(450,388){\usebox{\plotpoint}}
\put(449,390){\usebox{\plotpoint}}
\put(448,392){\usebox{\plotpoint}}
\put(447,394){\rule[-0.350pt]{0.700pt}{0.723pt}}
\put(446,397){\rule[-0.350pt]{0.700pt}{0.964pt}}
\put(445,401){\usebox{\plotpoint}}
\put(444,403){\usebox{\plotpoint}}
\put(443,405){\usebox{\plotpoint}}
\put(442,407){\usebox{\plotpoint}}
\put(441,409){\usebox{\plotpoint}}
\put(440,411){\rule[-0.350pt]{0.700pt}{0.964pt}}
\put(439,415){\usebox{\plotpoint}}
\put(438,417){\usebox{\plotpoint}}
\put(437,419){\usebox{\plotpoint}}
\put(436,421){\rule[-0.350pt]{0.700pt}{0.964pt}}
\put(435,425){\usebox{\plotpoint}}
\put(434,427){\usebox{\plotpoint}}
\put(433,429){\rule[-0.350pt]{0.700pt}{0.964pt}}
\put(432,433){\usebox{\plotpoint}}
\put(431,435){\usebox{\plotpoint}}
\put(430,437){\rule[-0.350pt]{0.700pt}{1.204pt}}
\put(429,442){\usebox{\plotpoint}}
\put(428,444){\rule[-0.350pt]{0.700pt}{0.964pt}}
\put(427,448){\usebox{\plotpoint}}
\put(426,450){\rule[-0.350pt]{0.700pt}{0.964pt}}
\put(425,454){\usebox{\plotpoint}}
\put(424,456){\rule[-0.350pt]{0.700pt}{0.964pt}}
\put(423,460){\usebox{\plotpoint}}
\put(422,462){\rule[-0.350pt]{0.700pt}{0.964pt}}
\put(421,466){\rule[-0.350pt]{0.700pt}{0.964pt}}
\put(420,470){\usebox{\plotpoint}}
\put(419,472){\rule[-0.350pt]{0.700pt}{0.964pt}}
\put(418,476){\rule[-0.350pt]{0.700pt}{0.964pt}}
\put(417,480){\rule[-0.350pt]{0.700pt}{1.204pt}}
\put(416,485){\usebox{\plotpoint}}
\put(415,487){\rule[-0.350pt]{0.700pt}{0.964pt}}
\put(414,491){\rule[-0.350pt]{0.700pt}{0.964pt}}
\put(413,495){\rule[-0.350pt]{0.700pt}{0.964pt}}
\put(412,499){\rule[-0.350pt]{0.700pt}{0.964pt}}
\put(411,503){\rule[-0.350pt]{0.700pt}{0.964pt}}
\put(410,507){\rule[-0.350pt]{0.700pt}{0.964pt}}
\put(409,511){\rule[-0.350pt]{0.700pt}{0.964pt}}
\put(408,515){\rule[-0.350pt]{0.700pt}{0.964pt}}
\put(407,519){\rule[-0.350pt]{0.700pt}{1.686pt}}
\put(406,526){\rule[-0.350pt]{0.700pt}{0.964pt}}
\put(405,530){\rule[-0.350pt]{0.700pt}{0.964pt}}
\put(404,534){\rule[-0.350pt]{0.700pt}{1.445pt}}
\put(403,540){\rule[-0.350pt]{0.700pt}{0.964pt}}
\put(402,544){\rule[-0.350pt]{0.700pt}{0.723pt}}
\put(311,167){\makebox(0,0){$\star$}}
\put(324,167){\makebox(0,0){$\star$}}
\put(338,167){\makebox(0,0){$\star$}}
\put(351,167){\makebox(0,0){$\star$}}
\put(364,167){\makebox(0,0){$\star$}}
\put(377,167){\makebox(0,0){$\star$}}
\put(390,167){\makebox(0,0){$\star$}}
\put(315,186){\makebox(0,0){$\star$}}
\put(328,186){\makebox(0,0){$\star$}}
\put(341,186){\makebox(0,0){$\star$}}
\put(355,186){\makebox(0,0){$\star$}}
\put(368,186){\makebox(0,0){$\star$}}
\put(381,186){\makebox(0,0){$\star$}}
\put(394,186){\makebox(0,0){$\star$}}
\put(319,204){\makebox(0,0){$\star$}}
\put(332,204){\makebox(0,0){$\star$}}
\put(345,204){\makebox(0,0){$\star$}}
\put(359,204){\makebox(0,0){$\star$}}
\put(372,204){\makebox(0,0){$\star$}}
\put(385,204){\makebox(0,0){$\star$}}
\put(399,204){\makebox(0,0){$\star$}}
\put(323,223){\makebox(0,0){$\star$}}
\put(336,223){\makebox(0,0){$\star$}}
\put(350,223){\makebox(0,0){$\star$}}
\put(363,223){\makebox(0,0){$\star$}}
\put(377,223){\makebox(0,0){$\star$}}
\put(390,223){\makebox(0,0){$\star$}}
\put(403,223){\makebox(0,0){$\star$}}
\put(327,241){\makebox(0,0){$\star$}}
\put(341,241){\makebox(0,0){$\star$}}
\put(354,241){\makebox(0,0){$\star$}}
\put(368,241){\makebox(0,0){$\star$}}
\put(381,241){\makebox(0,0){$\star$}}
\put(395,241){\makebox(0,0){$\star$}}
\put(408,241){\makebox(0,0){$\star$}}
\put(332,260){\makebox(0,0){$\star$}}
\put(346,260){\makebox(0,0){$\star$}}
\put(359,260){\makebox(0,0){$\star$}}
\put(373,260){\makebox(0,0){$\star$}}
\put(387,260){\makebox(0,0){$\star$}}
\put(400,260){\makebox(0,0){$\star$}}
\put(414,260){\makebox(0,0){$\star$}}
\put(337,278){\makebox(0,0){$\star$}}
\put(351,278){\makebox(0,0){$\star$}}
\put(365,278){\makebox(0,0){$\star$}}
\put(378,278){\makebox(0,0){$\star$}}
\put(392,278){\makebox(0,0){$\star$}}
\put(406,278){\makebox(0,0){$\star$}}
\put(420,278){\makebox(0,0){$\star$}}
\put(341,297){\makebox(0,0){$\star$}}
\put(355,297){\makebox(0,0){$\star$}}
\put(370,297){\makebox(0,0){$\star$}}
\put(384,297){\makebox(0,0){$\star$}}
\put(398,297){\makebox(0,0){$\star$}}
\put(413,297){\makebox(0,0){$\star$}}
\put(427,297){\makebox(0,0){$\star$}}
\put(344,315){\makebox(0,0){$\star$}}
\put(359,315){\makebox(0,0){$\star$}}
\put(374,315){\makebox(0,0){$\star$}}
\put(389,315){\makebox(0,0){$\star$}}
\put(404,315){\makebox(0,0){$\star$}}
\put(419,315){\makebox(0,0){$\star$}}
\put(434,315){\makebox(0,0){$\star$}}
\put(347,334){\makebox(0,0){$\star$}}
\put(363,334){\makebox(0,0){$\star$}}
\put(379,334){\makebox(0,0){$\star$}}
\put(395,334){\makebox(0,0){$\star$}}
\put(411,334){\makebox(0,0){$\star$}}
\put(427,334){\makebox(0,0){$\star$}}
\put(442,334){\makebox(0,0){$\star$}}
\put(349,353){\makebox(0,0){$\star$}}
\put(366,353){\makebox(0,0){$\star$}}
\put(383,353){\makebox(0,0){$\star$}}
\put(400,353){\makebox(0,0){$\star$}}
\put(417,353){\makebox(0,0){$\star$}}
\put(434,353){\makebox(0,0){$\star$}}
\put(451,353){\makebox(0,0){$\star$}}
\put(468,353){\makebox(0,0){$\star$}}
\put(485,353){\makebox(0,0){$\star$}}
\put(502,353){\makebox(0,0){$\star$}}
\put(519,353){\makebox(0,0){$\star$}}
\put(536,353){\makebox(0,0){$\star$}}
\put(347,371){\makebox(0,0){$\star$}}
\put(363,371){\makebox(0,0){$\star$}}
\put(379,371){\makebox(0,0){$\star$}}
\put(395,371){\makebox(0,0){$\star$}}
\put(411,371){\makebox(0,0){$\star$}}
\put(427,371){\makebox(0,0){$\star$}}
\put(442,371){\makebox(0,0){$\star$}}
\put(344,390){\makebox(0,0){$\star$}}
\put(359,390){\makebox(0,0){$\star$}}
\put(374,390){\makebox(0,0){$\star$}}
\put(389,390){\makebox(0,0){$\star$}}
\put(404,390){\makebox(0,0){$\star$}}
\put(419,390){\makebox(0,0){$\star$}}
\put(434,390){\makebox(0,0){$\star$}}
\put(341,408){\makebox(0,0){$\star$}}
\put(355,408){\makebox(0,0){$\star$}}
\put(370,408){\makebox(0,0){$\star$}}
\put(384,408){\makebox(0,0){$\star$}}
\put(398,408){\makebox(0,0){$\star$}}
\put(413,408){\makebox(0,0){$\star$}}
\put(427,408){\makebox(0,0){$\star$}}
\put(337,427){\makebox(0,0){$\star$}}
\put(351,427){\makebox(0,0){$\star$}}
\put(365,427){\makebox(0,0){$\star$}}
\put(378,427){\makebox(0,0){$\star$}}
\put(392,427){\makebox(0,0){$\star$}}
\put(406,427){\makebox(0,0){$\star$}}
\put(420,427){\makebox(0,0){$\star$}}
\put(332,445){\makebox(0,0){$\star$}}
\put(346,445){\makebox(0,0){$\star$}}
\put(359,445){\makebox(0,0){$\star$}}
\put(373,445){\makebox(0,0){$\star$}}
\put(387,445){\makebox(0,0){$\star$}}
\put(400,445){\makebox(0,0){$\star$}}
\put(414,445){\makebox(0,0){$\star$}}
\put(327,464){\makebox(0,0){$\star$}}
\put(341,464){\makebox(0,0){$\star$}}
\put(354,464){\makebox(0,0){$\star$}}
\put(368,464){\makebox(0,0){$\star$}}
\put(381,464){\makebox(0,0){$\star$}}
\put(395,464){\makebox(0,0){$\star$}}
\put(408,464){\makebox(0,0){$\star$}}
\put(323,482){\makebox(0,0){$\star$}}
\put(336,482){\makebox(0,0){$\star$}}
\put(350,482){\makebox(0,0){$\star$}}
\put(363,482){\makebox(0,0){$\star$}}
\put(377,482){\makebox(0,0){$\star$}}
\put(390,482){\makebox(0,0){$\star$}}
\put(403,482){\makebox(0,0){$\star$}}
\put(319,501){\makebox(0,0){$\star$}}
\put(332,501){\makebox(0,0){$\star$}}
\put(345,501){\makebox(0,0){$\star$}}
\put(359,501){\makebox(0,0){$\star$}}
\put(372,501){\makebox(0,0){$\star$}}
\put(385,501){\makebox(0,0){$\star$}}
\put(399,501){\makebox(0,0){$\star$}}
\put(315,519){\makebox(0,0){$\star$}}
\put(328,519){\makebox(0,0){$\star$}}
\put(341,519){\makebox(0,0){$\star$}}
\put(355,519){\makebox(0,0){$\star$}}
\put(368,519){\makebox(0,0){$\star$}}
\put(381,519){\makebox(0,0){$\star$}}
\put(394,519){\makebox(0,0){$\star$}}
\put(311,538){\makebox(0,0){$\star$}}
\put(324,538){\makebox(0,0){$\star$}}
\put(338,538){\makebox(0,0){$\star$}}
\put(351,538){\makebox(0,0){$\star$}}
\put(364,538){\makebox(0,0){$\star$}}
\put(377,538){\makebox(0,0){$\star$}}
\put(390,538){\makebox(0,0){$\star$}}
\end{picture}
\hspace{-2.5cm} 
\setlength{\unitlength}{0.240900pt}
\ifx\plotpoint\undefined\newsavebox{\plotpoint}\fi
\sbox{\plotpoint}{\rule[-0.175pt]{0.350pt}{0.350pt}}%
\begin{picture}(719,660)(0,0)
\tenrm
\sbox{\plotpoint}{\rule[-0.175pt]{0.350pt}{0.350pt}}%
\put(264,223){\rule[-0.175pt]{4.818pt}{0.350pt}}
\put(635,223){\rule[-0.175pt]{4.818pt}{0.350pt}}
\put(264,353){\rule[-0.175pt]{4.818pt}{0.350pt}}
\put(635,353){\rule[-0.175pt]{4.818pt}{0.350pt}}
\put(264,482){\rule[-0.175pt]{4.818pt}{0.350pt}}
\put(635,482){\rule[-0.175pt]{4.818pt}{0.350pt}}
\put(264,158){\rule[-0.175pt]{0.350pt}{4.818pt}}
\put(264,527){\rule[-0.175pt]{0.350pt}{4.818pt}}
\put(525,158){\rule[-0.175pt]{0.350pt}{4.818pt}}
\put(525,527){\rule[-0.175pt]{0.350pt}{4.818pt}}
\put(264,158){\rule[-0.175pt]{94.192pt}{0.350pt}}
\put(655,158){\rule[-0.175pt]{0.350pt}{93.710pt}}
\put(264,547){\rule[-0.175pt]{94.192pt}{0.350pt}}
\put(415,113){\makebox(0,0){$2${\large$zt$}$/B$}}
\put(368,469){\makebox(0,0)[l]{\small $n=2$}}
\put(400,223){\makebox(0,0)[l]{\small $n=4$}}
\put(546,184){\makebox(0,0)[l]{\small $n=6$}}
\put(561,504){\makebox(0,0)[l]{\LARGE (d)}}
\put(264,139){\makebox(0,0)[l]{\small 0}}
\put(514,139){\makebox(0,0)[l]{\small 0.5}}
\put(228,223){\makebox(0,0)[l]{\small -1}}
\put(238,353){\makebox(0,0)[l]{\small 0}}
\put(238,482){\makebox(0,0)[l]{\small 1}}
\put(264,158){\rule[-0.175pt]{0.350pt}{93.710pt}}
\sbox{\plotpoint}{\rule[-0.500pt]{1.000pt}{1.000pt}}%
\put(295,158){\usebox{\plotpoint}}
\put(296,160){\usebox{\plotpoint}}
\put(297,162){\usebox{\plotpoint}}
\put(298,164){\usebox{\plotpoint}}
\put(299,167){\usebox{\plotpoint}}
\put(300,170){\usebox{\plotpoint}}
\put(301,173){\usebox{\plotpoint}}
\put(302,176){\usebox{\plotpoint}}
\put(303,179){\usebox{\plotpoint}}
\put(304,181){\usebox{\plotpoint}}
\put(305,183){\usebox{\plotpoint}}
\put(306,185){\usebox{\plotpoint}}
\put(307,187){\usebox{\plotpoint}}
\put(308,189){\usebox{\plotpoint}}
\put(309,192){\usebox{\plotpoint}}
\put(310,194){\usebox{\plotpoint}}
\put(311,197){\usebox{\plotpoint}}
\put(312,198){\usebox{\plotpoint}}
\put(313,200){\usebox{\plotpoint}}
\put(314,202){\usebox{\plotpoint}}
\put(315,204){\usebox{\plotpoint}}
\put(316,206){\usebox{\plotpoint}}
\put(317,207){\usebox{\plotpoint}}
\put(318,209){\usebox{\plotpoint}}
\put(319,211){\usebox{\plotpoint}}
\put(320,212){\usebox{\plotpoint}}
\put(321,214){\usebox{\plotpoint}}
\put(322,216){\usebox{\plotpoint}}
\put(323,217){\usebox{\plotpoint}}
\put(324,218){\usebox{\plotpoint}}
\put(325,220){\usebox{\plotpoint}}
\put(326,221){\usebox{\plotpoint}}
\put(327,223){\usebox{\plotpoint}}
\put(328,224){\usebox{\plotpoint}}
\put(329,226){\usebox{\plotpoint}}
\put(330,227){\usebox{\plotpoint}}
\put(331,228){\usebox{\plotpoint}}
\put(332,230){\usebox{\plotpoint}}
\put(333,231){\usebox{\plotpoint}}
\put(334,233){\usebox{\plotpoint}}
\put(335,234){\usebox{\plotpoint}}
\put(336,236){\usebox{\plotpoint}}
\put(337,237){\usebox{\plotpoint}}
\put(338,238){\usebox{\plotpoint}}
\put(339,239){\usebox{\plotpoint}}
\put(340,241){\usebox{\plotpoint}}
\put(341,242){\usebox{\plotpoint}}
\put(342,243){\usebox{\plotpoint}}
\put(343,245){\usebox{\plotpoint}}
\put(344,246){\usebox{\plotpoint}}
\put(345,247){\usebox{\plotpoint}}
\put(346,248){\usebox{\plotpoint}}
\put(347,249){\usebox{\plotpoint}}
\put(348,250){\usebox{\plotpoint}}
\put(349,251){\usebox{\plotpoint}}
\put(350,252){\usebox{\plotpoint}}
\put(351,253){\usebox{\plotpoint}}
\put(352,255){\usebox{\plotpoint}}
\put(353,256){\usebox{\plotpoint}}
\put(354,257){\usebox{\plotpoint}}
\put(355,258){\usebox{\plotpoint}}
\put(356,259){\usebox{\plotpoint}}
\put(357,260){\usebox{\plotpoint}}
\put(358,261){\usebox{\plotpoint}}
\put(359,262){\usebox{\plotpoint}}
\put(360,263){\usebox{\plotpoint}}
\put(361,265){\usebox{\plotpoint}}
\put(362,266){\usebox{\plotpoint}}
\put(363,267){\usebox{\plotpoint}}
\put(364,268){\usebox{\plotpoint}}
\put(365,269){\usebox{\plotpoint}}
\put(366,270){\usebox{\plotpoint}}
\put(367,271){\usebox{\plotpoint}}
\put(368,272){\usebox{\plotpoint}}
\put(369,273){\usebox{\plotpoint}}
\put(370,275){\usebox{\plotpoint}}
\put(371,276){\usebox{\plotpoint}}
\put(372,277){\usebox{\plotpoint}}
\put(374,278){\usebox{\plotpoint}}
\put(375,279){\usebox{\plotpoint}}
\put(376,280){\usebox{\plotpoint}}
\put(378,281){\usebox{\plotpoint}}
\put(379,282){\usebox{\plotpoint}}
\put(380,283){\usebox{\plotpoint}}
\put(382,284){\usebox{\plotpoint}}
\put(383,285){\usebox{\plotpoint}}
\put(384,286){\usebox{\plotpoint}}
\put(385,287){\usebox{\plotpoint}}
\put(387,288){\usebox{\plotpoint}}
\put(388,289){\usebox{\plotpoint}}
\put(389,290){\usebox{\plotpoint}}
\put(391,291){\usebox{\plotpoint}}
\put(392,292){\usebox{\plotpoint}}
\put(393,293){\usebox{\plotpoint}}
\put(394,294){\usebox{\plotpoint}}
\put(396,295){\usebox{\plotpoint}}
\put(398,296){\usebox{\plotpoint}}
\put(399,297){\usebox{\plotpoint}}
\put(401,298){\usebox{\plotpoint}}
\put(402,299){\usebox{\plotpoint}}
\put(404,300){\usebox{\plotpoint}}
\put(405,301){\usebox{\plotpoint}}
\put(407,302){\usebox{\plotpoint}}
\put(408,303){\usebox{\plotpoint}}
\put(410,304){\usebox{\plotpoint}}
\put(411,305){\usebox{\plotpoint}}
\put(413,306){\usebox{\plotpoint}}
\put(415,307){\usebox{\plotpoint}}
\put(417,308){\usebox{\plotpoint}}
\put(418,309){\usebox{\plotpoint}}
\put(420,310){\usebox{\plotpoint}}
\put(422,311){\usebox{\plotpoint}}
\put(424,312){\usebox{\plotpoint}}
\put(426,313){\usebox{\plotpoint}}
\put(427,314){\usebox{\plotpoint}}
\put(430,315){\usebox{\plotpoint}}
\put(432,316){\usebox{\plotpoint}}
\put(435,317){\usebox{\plotpoint}}
\put(437,318){\usebox{\plotpoint}}
\put(439,319){\usebox{\plotpoint}}
\put(442,320){\usebox{\plotpoint}}
\put(444,321){\usebox{\plotpoint}}
\put(446,322){\usebox{\plotpoint}}
\put(449,323){\usebox{\plotpoint}}
\put(451,324){\usebox{\plotpoint}}
\put(453,325){\usebox{\plotpoint}}
\put(456,326){\usebox{\plotpoint}}
\put(458,327){\usebox{\plotpoint}}
\put(460,328){\usebox{\plotpoint}}
\put(463,329){\usebox{\plotpoint}}
\put(465,330){\usebox{\plotpoint}}
\put(468,331){\usebox{\plotpoint}}
\put(470,332){\usebox{\plotpoint}}
\put(472,333){\usebox{\plotpoint}}
\put(475,334){\usebox{\plotpoint}}
\put(477,335){\usebox{\plotpoint}}
\put(479,336){\usebox{\plotpoint}}
\put(481,337){\usebox{\plotpoint}}
\put(484,338){\usebox{\plotpoint}}
\put(486,339){\usebox{\plotpoint}}
\put(488,340){\usebox{\plotpoint}}
\put(490,341){\usebox{\plotpoint}}
\put(492,342){\usebox{\plotpoint}}
\put(495,343){\usebox{\plotpoint}}
\put(498,344){\usebox{\plotpoint}}
\put(501,345){\usebox{\plotpoint}}
\put(504,346){\usebox{\plotpoint}}
\put(507,347){\usebox{\plotpoint}}
\put(510,348){\usebox{\plotpoint}}
\put(513,349){\usebox{\plotpoint}}
\put(516,350){\usebox{\plotpoint}}
\put(519,351){\usebox{\plotpoint}}
\put(522,352){\usebox{\plotpoint}}
\put(525,353){\rule[-0.500pt]{31.317pt}{1.000pt}}
\put(521,353){\rule[-0.500pt]{32.120pt}{1.000pt}}
\put(518,354){\usebox{\plotpoint}}
\put(515,355){\usebox{\plotpoint}}
\put(511,356){\usebox{\plotpoint}}
\put(508,357){\usebox{\plotpoint}}
\put(505,358){\usebox{\plotpoint}}
\put(501,359){\usebox{\plotpoint}}
\put(498,360){\usebox{\plotpoint}}
\put(495,361){\usebox{\plotpoint}}
\put(492,362){\usebox{\plotpoint}}
\put(490,363){\usebox{\plotpoint}}
\put(488,364){\usebox{\plotpoint}}
\put(486,365){\usebox{\plotpoint}}
\put(483,366){\usebox{\plotpoint}}
\put(481,367){\usebox{\plotpoint}}
\put(479,368){\usebox{\plotpoint}}
\put(477,369){\usebox{\plotpoint}}
\put(475,370){\usebox{\plotpoint}}
\put(473,371){\usebox{\plotpoint}}
\put(470,372){\usebox{\plotpoint}}
\put(468,373){\usebox{\plotpoint}}
\put(465,374){\usebox{\plotpoint}}
\put(463,375){\usebox{\plotpoint}}
\put(461,376){\usebox{\plotpoint}}
\put(458,377){\usebox{\plotpoint}}
\put(456,378){\usebox{\plotpoint}}
\put(453,379){\usebox{\plotpoint}}
\put(451,380){\usebox{\plotpoint}}
\put(449,381){\usebox{\plotpoint}}
\put(446,382){\usebox{\plotpoint}}
\put(444,383){\usebox{\plotpoint}}
\put(441,384){\usebox{\plotpoint}}
\put(439,385){\usebox{\plotpoint}}
\put(437,386){\usebox{\plotpoint}}
\put(434,387){\usebox{\plotpoint}}
\put(432,388){\usebox{\plotpoint}}
\put(430,389){\usebox{\plotpoint}}
\put(428,390){\usebox{\plotpoint}}
\put(426,391){\usebox{\plotpoint}}
\put(424,392){\usebox{\plotpoint}}
\put(422,393){\usebox{\plotpoint}}
\put(420,394){\usebox{\plotpoint}}
\put(419,395){\usebox{\plotpoint}}
\put(417,396){\usebox{\plotpoint}}
\put(415,397){\usebox{\plotpoint}}
\put(413,398){\usebox{\plotpoint}}
\put(411,399){\usebox{\plotpoint}}
\put(410,400){\usebox{\plotpoint}}
\put(408,401){\usebox{\plotpoint}}
\put(407,402){\usebox{\plotpoint}}
\put(405,403){\usebox{\plotpoint}}
\put(404,404){\usebox{\plotpoint}}
\put(402,405){\usebox{\plotpoint}}
\put(401,406){\usebox{\plotpoint}}
\put(399,407){\usebox{\plotpoint}}
\put(398,408){\usebox{\plotpoint}}
\put(396,409){\usebox{\plotpoint}}
\put(395,410){\usebox{\plotpoint}}
\put(393,411){\usebox{\plotpoint}}
\put(392,412){\usebox{\plotpoint}}
\put(391,413){\usebox{\plotpoint}}
\put(389,414){\usebox{\plotpoint}}
\put(388,415){\usebox{\plotpoint}}
\put(387,416){\usebox{\plotpoint}}
\put(385,417){\usebox{\plotpoint}}
\put(384,418){\usebox{\plotpoint}}
\put(383,419){\usebox{\plotpoint}}
\put(382,420){\usebox{\plotpoint}}
\put(380,421){\usebox{\plotpoint}}
\put(379,422){\usebox{\plotpoint}}
\put(377,423){\usebox{\plotpoint}}
\put(376,424){\usebox{\plotpoint}}
\put(375,425){\usebox{\plotpoint}}
\put(373,426){\usebox{\plotpoint}}
\put(372,427){\usebox{\plotpoint}}
\put(371,428){\usebox{\plotpoint}}
\put(370,429){\usebox{\plotpoint}}
\put(370,430){\usebox{\plotpoint}}
\put(369,431){\usebox{\plotpoint}}
\put(368,432){\usebox{\plotpoint}}
\put(367,433){\usebox{\plotpoint}}
\put(366,434){\usebox{\plotpoint}}
\put(365,435){\usebox{\plotpoint}}
\put(364,436){\usebox{\plotpoint}}
\put(363,437){\usebox{\plotpoint}}
\put(362,438){\usebox{\plotpoint}}
\put(361,440){\usebox{\plotpoint}}
\put(360,441){\usebox{\plotpoint}}
\put(359,442){\usebox{\plotpoint}}
\put(358,443){\usebox{\plotpoint}}
\put(357,444){\usebox{\plotpoint}}
\put(356,445){\usebox{\plotpoint}}
\put(355,446){\usebox{\plotpoint}}
\put(354,447){\usebox{\plotpoint}}
\put(353,448){\usebox{\plotpoint}}
\put(352,450){\usebox{\plotpoint}}
\put(351,451){\usebox{\plotpoint}}
\put(350,452){\usebox{\plotpoint}}
\put(349,453){\usebox{\plotpoint}}
\put(348,454){\usebox{\plotpoint}}
\put(347,455){\usebox{\plotpoint}}
\put(346,456){\usebox{\plotpoint}}
\put(345,457){\usebox{\plotpoint}}
\put(344,458){\usebox{\plotpoint}}
\put(343,460){\usebox{\plotpoint}}
\put(342,461){\usebox{\plotpoint}}
\put(341,462){\usebox{\plotpoint}}
\put(340,463){\usebox{\plotpoint}}
\put(339,465){\usebox{\plotpoint}}
\put(338,466){\usebox{\plotpoint}}
\put(337,467){\usebox{\plotpoint}}
\put(336,468){\usebox{\plotpoint}}
\put(335,470){\usebox{\plotpoint}}
\put(334,471){\usebox{\plotpoint}}
\put(333,473){\usebox{\plotpoint}}
\put(332,474){\usebox{\plotpoint}}
\put(331,476){\usebox{\plotpoint}}
\put(330,477){\usebox{\plotpoint}}
\put(329,478){\usebox{\plotpoint}}
\put(328,480){\usebox{\plotpoint}}
\put(327,481){\usebox{\plotpoint}}
\put(326,483){\usebox{\plotpoint}}
\put(325,484){\usebox{\plotpoint}}
\put(324,486){\usebox{\plotpoint}}
\put(323,487){\usebox{\plotpoint}}
\put(322,488){\usebox{\plotpoint}}
\put(321,490){\usebox{\plotpoint}}
\put(320,492){\usebox{\plotpoint}}
\put(319,493){\usebox{\plotpoint}}
\put(318,495){\usebox{\plotpoint}}
\put(317,497){\usebox{\plotpoint}}
\put(316,498){\usebox{\plotpoint}}
\put(315,500){\usebox{\plotpoint}}
\put(314,502){\usebox{\plotpoint}}
\put(313,504){\usebox{\plotpoint}}
\put(312,506){\usebox{\plotpoint}}
\put(311,507){\usebox{\plotpoint}}
\put(310,510){\usebox{\plotpoint}}
\put(309,513){\usebox{\plotpoint}}
\put(308,515){\usebox{\plotpoint}}
\put(307,518){\usebox{\plotpoint}}
\put(306,520){\usebox{\plotpoint}}
\put(305,522){\usebox{\plotpoint}}
\put(304,524){\usebox{\plotpoint}}
\put(303,526){\usebox{\plotpoint}}
\put(302,528){\usebox{\plotpoint}}
\put(301,531){\usebox{\plotpoint}}
\put(300,534){\usebox{\plotpoint}}
\put(299,537){\usebox{\plotpoint}}
\put(298,540){\usebox{\plotpoint}}
\put(297,542){\usebox{\plotpoint}}
\put(296,544){\usebox{\plotpoint}}
\put(504,158){\rule[-0.500pt]{1.000pt}{1.445pt}}
\put(505,164){\rule[-0.500pt]{1.000pt}{1.494pt}}
\put(506,170){\rule[-0.500pt]{1.000pt}{1.494pt}}
\put(507,176){\rule[-0.500pt]{1.000pt}{1.494pt}}
\put(508,182){\rule[-0.500pt]{1.000pt}{1.494pt}}
\put(509,188){\rule[-0.500pt]{1.000pt}{1.494pt}}
\put(510,194){\rule[-0.500pt]{1.000pt}{1.542pt}}
\put(511,201){\rule[-0.500pt]{1.000pt}{1.542pt}}
\put(512,207){\rule[-0.500pt]{1.000pt}{1.542pt}}
\put(513,214){\rule[-0.500pt]{1.000pt}{1.542pt}}
\put(514,220){\rule[-0.500pt]{1.000pt}{1.542pt}}
\put(515,226){\rule[-0.500pt]{1.000pt}{2.489pt}}
\put(516,237){\rule[-0.500pt]{1.000pt}{2.489pt}}
\put(517,247){\rule[-0.500pt]{1.000pt}{2.489pt}}
\put(518,258){\rule[-0.500pt]{1.000pt}{2.570pt}}
\put(519,268){\rule[-0.500pt]{1.000pt}{2.570pt}}
\put(520,279){\rule[-0.500pt]{1.000pt}{2.570pt}}
\put(521,289){\rule[-0.500pt]{1.000pt}{3.734pt}}
\put(522,305){\rule[-0.500pt]{1.000pt}{3.734pt}}
\put(523,321){\rule[-0.500pt]{1.000pt}{3.854pt}}
\put(524,337){\rule[-0.500pt]{1.000pt}{3.854pt}}
\put(525,353){\rule[-0.500pt]{1.000pt}{3.734pt}}
\put(524,368){\rule[-0.500pt]{1.000pt}{3.734pt}}
\put(523,384){\rule[-0.500pt]{1.000pt}{3.734pt}}
\put(522,399){\rule[-0.500pt]{1.000pt}{3.734pt}}
\put(521,415){\rule[-0.500pt]{1.000pt}{2.570pt}}
\put(520,425){\rule[-0.500pt]{1.000pt}{2.570pt}}
\put(519,436){\rule[-0.500pt]{1.000pt}{2.570pt}}
\put(518,446){\rule[-0.500pt]{1.000pt}{2.489pt}}
\put(517,457){\rule[-0.500pt]{1.000pt}{2.489pt}}
\put(516,467){\rule[-0.500pt]{1.000pt}{2.489pt}}
\put(515,478){\rule[-0.500pt]{1.000pt}{1.542pt}}
\put(514,484){\rule[-0.500pt]{1.000pt}{1.542pt}}
\put(513,490){\rule[-0.500pt]{1.000pt}{1.542pt}}
\put(512,497){\rule[-0.500pt]{1.000pt}{1.542pt}}
\put(511,503){\rule[-0.500pt]{1.000pt}{1.542pt}}
\put(510,509){\rule[-0.500pt]{1.000pt}{1.494pt}}
\put(509,516){\rule[-0.500pt]{1.000pt}{1.494pt}}
\put(508,522){\rule[-0.500pt]{1.000pt}{1.494pt}}
\put(507,528){\rule[-0.500pt]{1.000pt}{1.494pt}}
\put(506,534){\rule[-0.500pt]{1.000pt}{1.494pt}}
\put(505,541){\rule[-0.500pt]{1.000pt}{1.445pt}}
\put(520,158){\rule[-0.500pt]{1.000pt}{8.913pt}}
\put(521,195){\rule[-0.500pt]{1.000pt}{9.636pt}}
\put(522,235){\rule[-0.500pt]{1.000pt}{9.395pt}}
\put(523,274){\rule[-0.500pt]{1.000pt}{9.395pt}}
\put(524,313){\rule[-0.500pt]{1.000pt}{9.636pt}}
\put(525,353){\rule[-0.500pt]{1.000pt}{9.395pt}}
\put(524,392){\rule[-0.500pt]{1.000pt}{9.395pt}}
\put(523,431){\rule[-0.500pt]{1.000pt}{9.395pt}}
\put(522,470){\rule[-0.500pt]{1.000pt}{9.636pt}}
\put(521,510){\rule[-0.500pt]{1.000pt}{8.913pt}}
\end{picture}
}
\vspace{-1.65cm}
\end{center}
\caption{Phase diagrams in the $C\! -\! t$ plane for different values of $n$, 
the number of minima of the hindering potential. (a), (b), and (c) are for
one dimension. (d) is obtained from a mean-field approximation,
dependent on the coordination $z$; the phase boundaries for n=2,4,6 are
indicated in the graph. In every diagram the phase at the left is a 
disordered phase; the ones at the right, up and down, are ordered phases; 
stars denote critical points.}
\label{fig:crit}
\vspace{-0pt}
\end{figure}

Information for the full phase diagram ($C\!\neq\! 0$) can again be obtained
from the corresponding classical $d\!+\!1$ model.\cite{jose} 
For $d\!=\!1$ the phase diagram is displayed in Fig. 3 (a)-(c). 
In addition to the phases discussed above, 
a third phase appears, an ordered phase corresponding to the locking of $\phi$
around one of the $n$ equivalent minima. For $n\!>\!4$ the critical region 
separating the ordered and disordered regions has a finite width; it
shrinks to a line for $n\!=\!4$. For $n\!=\!2$ the model belongs to the Ising
universality class ($C\!\neq\! 0$). Indeed, the Hamiltonian maps
into a Ising model in a transverse magnetic field for $C\!\gg\! B$.

A mean-field calculation has been performed to obtain quantitative
estimates. We approximate the interaction term
by $4 z t \langle \cos \phi \rangle \! \sum_i \cos \phi_i$, $z$ being the
coordination number and $\langle \cos \phi \rangle$ the order 
parameter.\cite{simanek} The phase diagram [Fig. 3(d)] is expected to be 
more accurate in higher dimensions. The critical regions disappear, 
with classical order-disorder phase transitions remaining. 
Higher $n$ increases the $C$ range of the disordered phase.

\vspace{-3pt}

\section*{Acknowledgments}
\vspace{-2pt}
We acknowledge discussions with F. Yndur\'ain 
and financial support of DGICYT (Spain) grant PB92-0169. 
KGS was supported by Wellesley College.

\end{document}